\documentclass[fleqn,usenatbib]{mnras}

\usepackage{newtxtext,newtxmath}

\usepackage[T1]{fontenc}
\usepackage{ae,aecompl}

\usepackage{graphicx}	
\usepackage{amsmath}	
\usepackage{amssymb}	
\usepackage{rotating}
\usepackage{lscape}
\usepackage{breqn}		
\usepackage{mathtools}
\usepackage{multirow}
\usepackage{tabularx}
\usepackage{hyperref}
\usepackage[normalem]{ulem}

\newcommand{\RomanNumeralCaps}[1]
    {\MakeUppercase{\romannumeral #1}}

\newcommand{\RevHighlight}[2]
    {#2}

\usepackage{etoolbox}
\makeatletter
\patchcmd\@combinedblfloats{\box\@outputbox}{\unvbox\@outputbox}{}{%
  \errmessage{\noexpand\@combinedblfloats could not be patched}%
}%
\makeatother

\title[STEELIIa]{Predicting fully self-consistent satellite richness, galaxy growth and starformation rates from the STastical sEmi-Empirical modeL \textsc{steel}.}

\author[]{
Philip J. Grylls,$^{1}$\thanks{E-mail: P.Grylls@soton.ac.uk}
F. Shankar,$^{1}$\thanks{E-mail: F.Shankar@soton.ac.uk}
J. Leja,$^{2}$
N. Menci,$^{3}$
B. Moster,$^{4}$
P. Behroozi,$^{5}$
L. Zanisi$^{1}$
\\
$^{1}$ Department for Physics and Astronomy, University of Southampton, Highfield SO171BJ, UK\\
$^{2}$ Harvard-Smithsonian Center for Astrophysics, 60 Garden St. Cambridge, MA 02138, USA\\
$^{3}$ INAF - Osservatorio Astronomico di Roma, via di Frascati 33, 00040 Monte Porzio Catone, Italy\\
$^{4}$ Universit\"ats-Sternwarte, Ludwig-Maximilians-Universit\"at M\"unchen, Scheinerstr. 1,81679 M\"uchen, Germany\\
$^{5}$ Department of Astronomy and Steward Observatory, University of Arizona, Tucson, AZ 85721, USA\\
}

\date{Accepted 2019 October 18. Received 2019 October 14; in original form 2019 May 2}

\pubyear{2019}

\begin{document}
\label{firstpage}
\pagerange{\pageref{firstpage}--\pageref{lastpage}}
\maketitle

\begin{abstract}
Observational systematics complicate comparisons with theoretical models limiting understanding of galaxy evolution. In particular, different empirical determinations of the stellar mass function imply distinct mappings between the galaxy and halo masses, leading to diverse galaxy evolutionary tracks. Using our state-of-the-art STatistical sEmi-Empirical modeL, \textsc{steel}, we show fully self-consistent models capable of generating galaxy growth histories that simultaneously and closely agree with the latest data on satellite richness and star-formation rates at multiple redshifts and environments. Central galaxy histories are generated using the central halo mass tracks from state-of-the-art statistical dark matter accretion histories coupled to abundance matching routines. We show that too flat high-mass slopes in the input stellar-mass-halo-mass relations as predicted by previous works, imply non-physical stellar mass growth histories weaker than those implied by satellite accretion alone. Our best-fit models reproduce the satellite distributions at the largest masses and highest redshifts probed, the latest data on star formation rates and its bi-modality in the local Universe, and the correct fraction of ellipticals. Our results are important to predict robust and self-consistent stellar-mass-halo-mass relations and to generate reliable galaxy mock catalogues for the next generations of extra-galactic surveys such as Euclid and LSST.  
\end{abstract}




\section{Introduction}
Galaxies are thought to grow and evolve through a combination of `in-situ' and `ex-situ' processes. In-situ processes such as star formation, are thought to be driven by the availability of cold gas in the galaxy. The reserve of cold gas ready to fuel star formation could be regulated by a number of internal and external processes, from stellar and active galactic nuclei feedback to host halo and/or morphological quenching \citep[e.g.,][]{Granato2004AHosts, Dekel2009ColdFormation, Lilly2013GASHALOS, Schawinski2014TheGalaxies} One important ex-situ channel affecting galaxy growth is satellite accretion. In particular, in very massive galaxies growth via satellite accretion has been claimed to become progressively more relevant \citep{DeLucia2006TheGalaxies,vanDokkum2010THE2, Shankar2013SizeUniverse, Shankar2015, Buchan2016, Groenewald2017TheGrowth, Matharu2019HSTMergers}. Central galaxies that reside at the centre of massive haloes thus provide a window into the different pathways that have contributed to the mass growth history of galaxies in the local universe. Exploring the way these galaxies build their mass can give insights into the stellar-mass-halo-mass (SMHM hereafter) relation, the efficiency of the satellite transport from the edge of the cluster to the centre, the balance of the major processes taking place on these satellites, the galaxy merger rate, and the star formation rate. The characteristic mass at which galaxies transition from being in-situ to ex-situ growth dominated has previously been found at $M_* \sim 10^{11} M_{\odot}$ \citep{Cattaneo2011HowMass, Bernardi2011EvidenceRelations, Shankar2013SizeUniverse}. 

Models of galaxy formation traditionally use the hierarchical growth of dark matter structure as the backbone for galaxy assembly. Hydrodynamical simulations co-evolve the dark matter and baryonic matter allowing for a simultaneous look at the assembly of both components \citep{McAlpine2015TheCatalogues,Vogelsberger2014IntroducingUniverse}. The latter technique, however, requires large computational resources. Less computationally intensive models such as traditional Semi-analytic and Semi-empirical models use dark matter merger trees from post-processing of dark matter simulations \citep{Guo2011FromCosmology, Shankar2013}. Dark matter merger trees visualize dark matter assembly as a central trunk and halo mergers happen where branches join. Semi-analytic models initialize gas at high redshift and use a number of physical assumptions and free parameters to tune to observations \citep{DeLucia2006TheGalaxies, Guo2011FromCosmology}. Semi-empirical models use a more direct approach initializing galaxy stellar mass in dark matter haloes most commonly through abundance matching, the association of galaxies to dark matter host haloes via relative abundances \citep{Hopkins2010MERGERSMATTER, Zavala2012, Moster2013, Shankar2014, Moster2018Emerge10}. Both Semi-Empirical and Semi-Analytic models follow the merging histories of the underling dark matter merger trees to track the in-situ and ex-situ buildup of galaxy mass. The work of \citet{Moster2018Emerge10}, for example, uses a semi-empirical model to associate the growth of the dark matter halo to the star formation rate of the host galaxy alongside the build-up of stellar mass from satellites accretion, further strengthening the connection between the dark matter host environment and the build-up of galactic stellar mass. 

\RevHighlight{There is a known discrepancy between the observed star formation rate and growth of the galaxy stellar mass function. Previous observations of star formation rate The observed star formation rate is too high to be consistent with the evolution of central galaxies \citep[e.g.][]{Leja2015ReconcilingFunction, Lapi2017StellarEquation} and causes runaway growth inconsistent with observed population statistics of satellite galaxies \citep{Grylls2019AClusters}.}
{An average measure of the growth of galaxies can be obtained by comparing stellar mass functions, the number densities of galaxies as a function of mass, over multiple epochs. Selecting galaxy populations with the same number density at each epoch, assuming that galaxies maintain rank ordering over cosmic time, allows an estimation of the average growth of galaxies to be made. An estimation of the star formation rate for each galaxy population can then be computed by taking the time derivative of the stellar mass growth. However, the star formation generated in this way is significantly lower than observational estimates of the star formation rate \citep[e.g.][]{Leja2015ReconcilingFunction, Lapi2017StellarEquation}. It is consequently found that if observed star formation rates are used in models, they cannot be reconciled with the stellar mass functions. In \cite{Grylls2019AClusters} it is shown this is also in effect in satellite galaxy distributions, where the predicted number of massive satellites is far too high if satellites evolve using observed star formation rates.}
This is a particular problem for semi-empirical models where one would ideally use the observed star formation rate as an input. To overcome the inconsistencies between observed star formation rates and model predictions it is possible to include starformation rates generated by the method above commonly referred to as a continuity star formation rate. Attributing the stellar mass growth to star formation in this way yields an upper limit to star formation rate that is consistent with the stellar mass function evolution by design.

\RevHighlight{To properly constrain the formation of a galaxy one must reproduce the galaxy environment, the distribution of satellites around the central galaxy, at all previous redshifts. Discrepancies with observations of the the high redshift environment will propagate unpredictably in the modelled in-situ growth to compensate for the ex-situ mass deficit/surplus from incorrectly calibrated progenitor distributions. Reproducing the number density and distribution of galaxies has consistently proven a challenge for many semi-analytic models \citep{Asquith2018CosmicModels}, mainly due to the numerous parameters used.} 
{To properly constrain the formation of a galaxy one must reproduce the galaxy environment, the distribution of satellites around the central galaxy, at all previous redshifts. Discrepancies with observations of the the high redshift environment will cause modelled satellite stellar mass accretion rates that are either too high or too low. To account for such deficit/surplus modelled in-situ growth must compensate though other modelling parameters to maintain the evolution of the stellar mass density. Such compensation could, for example, be of the form of suppressed/enhanced star formation rate or alternatively an any number of other physical modelling parameters. Reproducing the number density and distribution of galaxies has however proven a challenge for many semi-analytic models \citep[e.g.][]{Asquith2018CosmicModels}. Furthermore, where semi-analytic models have included more physics via an increased number of modelling parameters, this has led to degeneracies that obscure which are the essential physical processes governing galaxy formation \citep[e.g.][]{Lapi2011Herschel-atlasGalaxies,Gonzalez2011EVOLUTION4}.}
Semi-empirical models, due to the direct initialization of galaxies to haloes and smaller parameter spaces, fare better as they can by design provide more clarity as to which modelling assumptions and related parameters are necessary to fit observations. 

In our previous work we presented \textsc{steel} \citep[][hereafter referred to as Paper \RomanNumeralCaps{1}]{Grylls2019AClusters}, a STastical sEmi-Empirical modeL. The basis of \textsc{steel} was the shift from discrete merger trees in favour of statistical halo growths and merging histories, which enables to probe galaxy environment unbiased by volume and mass resolution. In Paper \RomanNumeralCaps{1} we presented a detailed study of the richness of the galaxy group and cluster environments in the local Universe. In this work we extend the analysis of satellite richness from \textsc{steel} to high redshifts comparing with a large galaxy cluster survey, and state-of-the-art hydrodynamical simulations. Having a clear and well-constrained picture of the building up of satellite population then allows \textsc{steel} to create more reliable merger histories for central galaxies across cosmic time. Using \textsc{steel}'s improved merger histories we are then able to constrain the ex-situ growth and, by extension, derive more reliable estimates of the in-situ growth and the implied star formation rates of central galaxies. 

The paper is laid out as follows: In Section \ref{sec:Data} we present the halo and stellar mass functions we use for our abundance matching as well as the high redshift clusters we adopt to constrain the model performance. In Section \ref{sec:Method} we discuss \textsc{steel}, most importantly we provide an overview of the statistical merging history in Section \ref{subsec:StatMergeHist}, the abundance matching in Section \ref{subsec:AbundanceMatching}, as well as updates made to the model described in Paper \RomanNumeralCaps{1}. We begin the results by testing the high redshift halo merger rate in Section \ref{subsec:HaloMergerRate}, in Section \ref{subsec:HighRedshiftClusters} we then present the high redshift satellite galaxy distribution results, and in Section \ref{subsec:CentMassAcc} the growth of our galaxy population via in-situ and ex-situ processes. We then discuss our results in a wider context in Section \ref{sec:Discussion} and summarize in Section \ref{sec:Conclusions}.

\section{Data}
\label{sec:Data}
In this section we first describe the simulations used for the halo mass functions (HMF), then the data used to create the stellar mass functions (SMF). 
Together, the HMF and SMF are used to create a SMHM relation, described in Section \ref{subsec:AbundanceMatching}, which defines the galaxy-halo connection, essential to \textsc{steel}. 
We then provide details of the cluster data we compare to at high redshifts.
All the data presented in this section are converted, wherever necessary, to a \citet{Chabrier2003GalacticFunction} stellar initial mass function (IMF). In this work we adopt the Planck cosmology with $(\Omega_m, \Omega_{\Lambda}, \Omega_{b}, h, n, \sigma_8) = (0.31, 0.69, 0.05, 0.68, 0.97, 0.82)$\footnote{We note that Planck's best-fit cosmology has slightly different parameters than those adopted in some of the observations used in this work, such as the stellar mass functions ($(\Omega_m, h = (0.30, 0.70)$). Correcting the stellar mass function volumes and luminosities to the same cosmology yields essentially indistinguishable results.} \citep{PlanckCollaboration2015PlanckParameters}. Halo masses are defined as virial masses in this cosmology, unless stated otherwise.

\subsection{Halo Mass Functions}
\label{subsec:HMF}
In this work we use the halo mass function from the simulations of \citet{Despali2016TheDefinitions}, generated and converted to appropriate units and cosmology using \textsc{colossus} \citep{Diemer2017COLOSSUS:Halos}. 
The halo mass function provides the number density of haloes in a given mass bin at a given redshift. 
We generate the substructure of subhaloes using the subhalo mass functions found in \citet{Jiang2016StatisticsFunctions}. 
The subhalo mass function provides the number density of subhaloes expected for a parent halo of a given mass. 
The (sub) halo mass functions used in this work are all calibrated against the Bolshoi Simulation \citep{Klypin2016}.

\subsection{Stellar Mass Functions}
\label{subsec:SMF}

In this work we use the stellar mass functions defined below, along with the halo mass function given above, to constrain the SMHM relationship. The latter in \textsc{steel} is constrained first at low redshift $z = 0.1$, using stellar mass functions from the Sloan Digital Sky Survey (\ref{subsub:SDSS}).
The evolution of the SMHM relation is then constrained to match the stellar mass function at higher redshifts ($z >0.1$).

\subsubsection{Low Redshift, $z = 0.1$}
\label{subsub:SDSS}
At low redshift we use the Sloan Digital Sky Survey Data Release 7 (SDSS-DR7) from \citet{Meert2015ASystematics}.
The data from the SDSS-DR7 spectroscopic sample \citep{Abazajian2009THESURVEY} contain $\sim 670,000$ galaxies fitted with a S\'ersic + exponential model \citep[PyMorph;][]{Meert2015ASystematics} with associated halo masses and central satellite classifications from \citep{Yang2012EvolutionHalos}. The improved photometric analysis by \citet{Meert2015ASystematics} provides more reliable estimates of the stellar mass function at the high mass end which appear more abundant than previous estimates \citep{Bernardi2016TheEvolution, Bernardi2017ComparingLight}.
In this work we investigate the effect of this enhanced high mass end on galaxy assembly. \RevHighlight{}{We compare to previous determinations of the stellar mass function using as an example the de Vaucoulers \citep{deVaucouleurs1948RecherchesExtragalactiques} based cmodel fits from SDSS \citep{Abazajian2009THESURVEY}. The latter definition of galaxy stellar mass has been extensively discussed not to be accurate, partially due to incorrect sky subtraction and adoption of non-ideal light profiles \citep{Bernardi2013TheProfile}. \citet{Bernardi2017ComparingLight} have clearly shown that the choice of light profile is not a simple matter of "semantics". The single or double S\`ersic models perform better in fitting the surface brightness of galaxies independently of the galactic environment \citep{Meert2015ASystematics}. The performance is thus not related to the inclusion of the intra-group or intra-cluster light in the fit \citep{Bernardi2017ComparingLight}.}

\subsubsection{High Redshift, $z > 0.1$}
\label{subsub:Davidzon}
\RevHighlight{At higher redshift (0.3 < z < 3.3) we use stellar mass functions from the COSMOS2015 catalogue \citep{Davidzon2017TheSnapshots}. Here masses are defined using spectral energy distribution fitting (SED), including ultra-deep infrared photometry. \citet{Davidzon2017TheSnapshots} use \citet{Bruzual2003Stellar2003} stellar population synthesis models which are corrected by +0.15 dex to roughly match the \citet{Mendel2014ASURVEY} mass to light ratios when coupled to S\'ersic + exponential photometry.}{At higher redshift (0.3 < z < 3.3) we use stellar mass functions from the COSMOS2015 catalogue \citep{Davidzon2017TheSnapshots}. Here masses are defined using spectral energy distribution fitting, including ultra-deep infrared photometry. \citet{Davidzon2017TheSnapshots} use \citet{Bruzual2003Stellar2003} stellar population synthesis models to estimate stellar masses. As SED fitting is notably different from light profile fitting, one cannot apply the same corrections as in \citet{Mendel2014ASURVEY}. 
Nevertheless, to match the mass-to-light ratios adopted by \citet{Mendel2014ASURVEY}, based on the \citet{Bell2003TheFunctions} mass-to-light ratios, we follow \citet{Bernardi2013TheProfile} and increase the \citet{Davidzon2017TheSnapshots} stellar masses, based on \citet{Bruzual2003Stellar2003}, by +0.15 dex. We note that the resulting z=0.37 stellar mass function after this correction is in remarkable good agreement with the z=0.1 stellar mass function by \citep{Bernardi2013TheProfile}. Our result also matches the findings by \citet{Bernardi2016TheEvolution}, who showed that, by making use of the BOSS sample, the stellar mass function shows negligible number density evolution up to z \textasciitilde 0.5.} 

\subsection{Clusters}
\label{subsec:Clusters}
\subsubsection{Cluster at z = 2.5, Wang+ 2016}
\label{subsubsec:Wang}
The highest redshift cluster we compare to is a $M_{vir} = 10^{13.7} M_{\odot}$ halo containing 15 galaxies with $M_* > 10^{10} M_{\odot}$ at a redshift of $z = 2.5$.
This cluster is reported in \citet{Wang2016DiscoveryZ=2.506}, and we provide a brief description of the observation and data here.
The cluster is observed using IRAM-NOEMA, VLT-KMOS, VLA, XMM-Newton and Chandra for the spectroscopic observation and redshift determination.  
The galaxy masses are determined assuming a \citet{Salpeter1955TheEvolution.} IMF, which we correct to a \citet{Chabrier2003GalacticFunction} IMF, by decreasing the stellar masses by 0.24 dex.
The halo mass ($M_{vir} \sim 10^{13.93} M_{\odot}$) of the cluster is estimated in three different ways, using the total X-ray luminosity, the velocity dispersion of its member galaxies above $M_* = 10^{10.76} M_{\odot}$, and the stellar richness of the cluster \footnote{We note the velocity dispersions and X-ray luminosity estimations give the cluster mass as $M_{vir} = 10^{13.73} M_{\odot}$ and the estimate given by mass richness is significantly higher $M_{vir} = 10^{14.6} M_{\odot}$, whilst we used the published average the lower cluster mass excluding the richness estimate is in as-good or better agreement with model results.}. Given this object was a targeted cluster, we cannot estimate the cosmic abundance (i.e, the number per cubic megaparsec). For analysis and comparison later in this work we assign this cluster an abundance of $N(> M_*=10^{13.93})=10^{-7.15}$ $[Mpc^-3]$ which is estimated by integrating the halo mass function in the limits [$10^{13.93}$, $\infty$], thus providing an upper limit to the number densities associated to clusters of this mass. 

\subsubsection{1959 Clusters at z = 0.7 - 1.0, Wen \& Han 2018}
\label{subsubsec:1959}
We compare to the cluster sample from \citet{Wen2018ARedshifts}, which contains 1959 clusters from SDSS-DR14 \citep{Abolfathi2017TheExperiment} and the WISE survey \citep{Wright2010THEPERFORMANCE}. The clusters are identified in the W1 band, and foreground objects are removed using the SDSS photometric data. The cluster mass and richness are estimated using the total W1 band luminosity within 1 Mpc of the central galaxy. As performed above, to each cluster we assign an upper limit to their abundances from the cumulative integration of the halo mass function.

\section{Method: \textsc{steel}}
\label{sec:Method}
{Our model \textsc{steel} is a STastical sEmi-Empirical modeL designed to investigate the satellites and subhaloes in groups and clusters. In brief, \textsc{steel} removes reliance on discrete dark matter simulations or halo merger trees, commonly used in galaxy simulations, in favour of using mass functions to create a `statistical dark matter accretion history' described in more detail in Section \ref{subsec:StatMergeHist}. This statistical history is then combined with semi-empirical techniques, such as abundance matching (Section \ref{subsec:AbundanceMatching}), to create average galaxy population statistics. Whilst a comprehensive description of \textsc{steel} can be found in Paper \RomanNumeralCaps{1}, we provide in this section highlights of \textsc{steel}, including any relevant updates. In this paper we have two objectives. Firstly, we investigate the impact of different SMHM relations on the accretion histories of galaxies. The second aim is to use our semi-empirical accretion histories and galaxy growth histories to derive the star formation rate/star formation histories of galaxies.}

\begin{figure*}
	\centering
	\includegraphics[width = \linewidth]{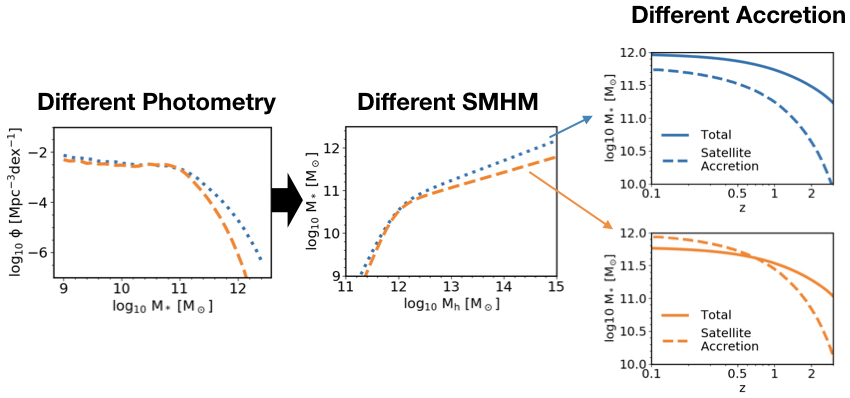}
    \caption{A cartoon showing the steps we follow to connect the differences found in the stellar mass function (left) and the changes in the SMHM relation (SMHM, middle), that propagate into changes in the accretion histories (right). In the right hand panel dashed lines are mass from satellite accretion and solid lines are total galaxy mass growth. Flatter SMHM relations imply a weaker growth of stellar mass in the central which can be easily overcome by the substantial cumulative growth of merging satellites, rendering the model internally inconsistent.}
	\label{fig:Simple_Cartoon}
\end{figure*}

\RevHighlight{The cartoon in Figure \ref{fig:Simple_Cartoon} shows a simple visualization of the process we use to determine the effect of photometry on the accretion histories of galaxy populations. From left to right; differences in the high mass end of the stellar mass function (left), propagate into a different slope of the SMHM relation (middle), the different slopes then give different ratios (right) of galaxy growth (solid lines) to galaxy accretion (dashed lines). The exact shape of the SMHM relation informs the ratio of accretion in two ways. Firstly, the low mass slope dictates the size of the satellite galaxies that merge with the central. Secondly, and most importantly for this work, the high mass slope affects the growth rate of the central galaxy for a given halo growth track a steeper slope implies more galaxy growth. Different combinations of these slopes change the satellite accretion ratios some of which may be nonphysical. For example, a slope that is too shallow the satellite accretion if far greater than the central galaxy growth. By checking the the ratio of accretion to central growth for each SMHM relation we provide an additional constraint to the stellar mass function in a given dark matter assembly.}{The cartoon in Figure \ref{fig:Simple_Cartoon} shows a simple visualization of the process we use to determine the effect of photometry on the accretion histories of galaxy populations. Starting on the left we show two stellar mass functions, the primary difference is the blue (dotted) stellar mass function has a substantially enhanced high mass end. In the middle panel we show how this high mass slope changes the SMHM relation, an enhanced high mass end stellar mass function results in an enhanced high mass slope. The galaxy growth histories, shown as solid lines, are generated using the SMHM relation used to calculate the average satellite stellar mass accretion associated to a given central halo mass history.
It follows that the galaxy grown using the steeper relation from the enhanced stellar mass function induces more galaxy growth. A flatter high mass slope induces less growth in the limit where the high mass slope is flat the central galaxy would not grow with increasing halo mass.}

\RevHighlight{}{The dashed lines in the rightmost panels are created by calculating the average accretion onto a central galaxy. The majority of accretion is from galaxy mergers that have a low mass ratio between the central and satellite galaxy. Note, the stellar mass functions show little difference in number density for smaller galaxies (below $\mathrm{M_*}$ = $10^{11}$ $\mathrm{M_{\odot}}$) and thus the low mass slope of the SMHM relation is unchanged. It follows that the smaller galaxies in a given dark matter assembly history are unchanged and the accretion is simmilar for both centrals. For the bottom right hand panel showing the accretion history and growth history for a galaxy using the lower stellar mass function it is found that the accretion exceeds the galaxy growth (in this cartoon this is accentuated for clarity). Whereas, for the enhanced SMHM relation the accretion is below that of the galaxy growth. The satellite galaxy accretion history and the central galaxy growth history in a given cosmology are determined by the SMHM relationship and the dark matter halo assembly. In this paper we describe a method that, in a given cosmology, can exclude a set of stellar mass functions over multiple redshift epochs. The evolution of these stellar mass functions is analogous with the growth of the total stellar mass in the Universe over cosmic time. The consistency of the galaxy growth histories and the satellite accretion histories is checked by generating and comparing the ratio of satellite accretion and total mass growth. If the total accretion or rate of accretion is greater than that of the central galaxy mass or galaxy growth rate the set of stellar mass functions is incompatible with the specific $\Lambda$CDM cosmology.}

\begin{figure*}
	\centering
	\includegraphics[width = \linewidth]{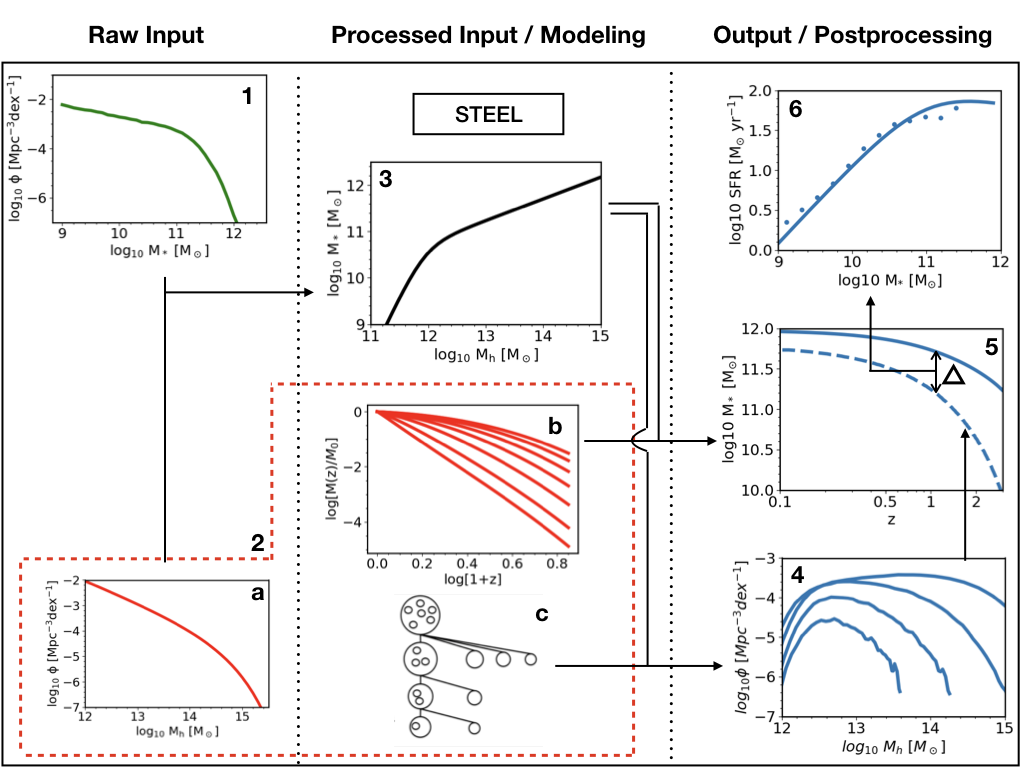}
    \caption{A cartoon showing the constituent steps of the method to generate star formation rates in this paper. In brief, the three columns from left to right are raw inputs, derived inputs/modelling, and output/post-processing. The subplots are: 1. The stellar mass function, 2a. The halo mass function, 2b. Halo mass growth histories, 2c. Accretion histories/Merger tree, 3. The SMHM relation, 4. Group/Cluster satellite richness, 5. Central growth histories/satellite accretion histories, 6. Star formation rate. The star formation rates are derived from the difference between the total growth in stellar mass and that from satellite accretion (panel 5).}
	\label{fig:Complex_Cartoon}
\end{figure*}

\RevHighlight{The cartoon in Figure \ref{fig:Complex_Cartoon} shows a simplification of the processes we follow to derive the star formation rate by following galaxy populations along their halo mass histories. From left to right we move from raw inputs in the left column, derived inputs and modelling in the middle column and outputs and post processing in the rightmost column. The plot labelled 1 (green) is the input stellar mass function. The box in red is the statistical dark matter accretion history described in Section \ref{subsec:StatMergeHist}, including the halo mass function (2a), the central growth histories (2b), and the substructure (2c) shown here as a discrete merger tree for visualization purposes. The stellar mass function (1) and the halo mass function (2a) are used to create the SMHM relationship (3, black) using the abundance matching routines described in Section \ref{subsec:AbundanceMatching}. \textsc{steel} then uses the SMHM relationship (2) in combination with the dark matter accretion histories (2c) to generate the distribution of satellite galaxies within central haloes at multiple redshifts (4), which is then tested against simulations and data in Section \ref{subsec:HighRedshiftClusters}. We can then use the satellite accretion inferred from the cluster richness (4) to generate average central galaxy accretion histories (dashed line, 5). Using the central halo growth histories (2b) and the SMHM relation (3), we can generate the average central galaxy growth history (solid line, 5). By comparing the deficit between the accreted mass and the growth history in 5 we can generate the star formation rate for central galaxies (solid line, 6) to compare to observational data (points, 6). Inserts in blue are results that are revisited in Sections \ref{sec:Results} and \ref{sec:Discussion}. We briefly summarize the methodology of \textsc{steel} in Section \ref{subsec:StatMergeHist} then provide the abundance matching methodology in Section \ref{subsec:AbundanceMatching}.}{The cartoon in Figure \ref{fig:Complex_Cartoon} shows a simplification of the processes we follow to derive the star formation rate by following galaxy populations along their halo mass histories. The plot labelled 1 (green) is the input stellar mass function. The box in red is the statistical dark matter accretion history described in Section \ref{subsec:StatMergeHist}, including the halo mass function (2a), the central growth histories (2b), and the halo substructure (2c) shown here as a discrete merger tree for visualization purposes. Using the abundance matching routines described in Section \ref{subsec:AbundanceMatching}, the stellar mass function (1) and the halo mass function (2a) are used to create the SMHM relationship (3, black). In Paper \RomanNumeralCaps{1} we showed how the dark matter accretion histories (2) and abundance matching (3) can be used to generate distributions of satellites for any central halo. In this work we generate satellite distributions for central haloes at multiple redshifts (4) then test them against simulations and observations in sections Section \ref{subsec:HighRedshiftClusters}. For each central halo mass track (2b) the average number density of satellites that reach the center of the halo and merge with the central galaxy, is calculated thus generating the average satellite accretion history (dashed line, 5). Using the central halo growth histories (2b) and the SMHM relation (3), we can generate the average central galaxy growth history (solid line, 5). These two quantities can be compared to check for self-consistency, as described above and shown in Figure \ref{fig:Simple_Cartoon}. Where a self-consistent central growth and accretion history is found any deficit between the accreted mass and the growth history is attributed to star formation rate (delta, 5). The derived star formation rate for central galaxies (solid line, 6) is compared to observational data (points, 6). Where the star formation rate prediction generated form the model is found to be consistent with the observed star formation rate this is a good indication that the model is correct. Additional observational constraints not shown in Figure \ref{fig:Complex_Cartoon} can be added to improve the analysis, such as the specific star formation rate distribution which is discussed in Section \ref{subsec:sSFR}. In future work other constraints such as the pair fraction of galaxies and the intracluster light generated from dynamical process during satellite accretion will also be considered.}

\subsection{Statistical Merging History}
\label{subsec:StatMergeHist}
Common modelling techniques, such as hydrodynamical, semi-analytic, and traditional semi empirical models rely on a discrete set of haloes within a simulation `volume'. These discrete haloes come in three forms: a N-body cosmological box, post-processed merger trees, or catalogue of haloes\footnote{We note that in many cases merger trees and halo catalogs are extracted from the cosmological box of a dark matter only N-body simulation. The alternatives being Press-Schecter analytically derived merger trees \citep[e.g.][]{Parkinson2008GeneratingTrees,Press1974} and halo catalogs obtained by sampling the halo mass function.}. Due to the inevitably restrained simulated cosmic volumes, all the aforementioned models are biased towards smaller haloes and galaxies, largely missing a statistically-comprehensive description of the most massive central and satellite galaxies. 

\textsc{steel} is instead designed to model all haloes (and subhaloes) within the simulation range without volume or resolution constraints. We remove the dependence on discrete halo sets through the use of a \textit{statistical} dark matter backbone, and simulate all haloes and all mass-ratio mergers\footnote{Within the simulation mass range, however, this can be set arbitrarily wide as long as the choice of halo mass function and empirical techniques are valid in the proposed mass range.} with equal weights regardless of their number density. The following steps represent the core method for creating the statistical dark matter accretion histories.

\begin{enumerate}
\item At the redshift of interest $\bar{z}$ we start from the halo mass function to compute the abundances of (parent/central) haloes for any given central halo bin $[M_{h,cent}(z),$ $M_{h,cent}(z) + dM_{h,cent}(z)]$. 
\item Each parent/central halo mass is then followed backwards in time following its average mass growth history, $<M_{\rm halo}(\bar{z})>$, calculated using the routines from \citet{vandenBosch2014ComingWells}.
\item An unevolved subhalo mass function\footnote{The unevolved subhalo mass function gives the total number density of subhaloes of each mass accreted onto a given central halo over it's entire growth history.} \citep{Jiang2016StatisticsFunctions} is then assigned to each central halo mass bin $[M_{h,cent}(z),$ $M_{h,cent}(z) + dM_{h,cent}(z)]$ at each redshift epoch.
\item At each time step we then calculate the difference in subhalo population
between $z$ and $z+dz$ to estimate the average number density and masses of subhaloes accreted onto the main progenitor in the redshift interval $dz$, which we call the `accreted' subhalo mass function.
\item Each bin $[M_{h,sub},$ $M_{h,sub} + dM_{h,sub}]$ of the accreted subhalo mass function is then assigned a dynamical time given the central halo mass bin $[M_{h,cent}(z),$ $M_{h,cent}(z) + dM_{h,cent}(z)]$ it corresponds to.
\item At each redshift epoch we then sum the number densities of subhalo bins $[M_{h,sub},$ $M_{h,sub} + dM_{h,sub}]$ on each central halo mass track that have not exceeded their dynamical time to create the surviving subhalo mass function.
\item Given the infall redshift, mass and number densities of each subhalo mass bin $[M_{h,sub},$ $M_{h,sub} + dM_{h,sub}]$ we can convolve the resulting satellite halo distribution with the SMHM relation to create the observed distribution of satellites at any epoch. In some model variants we also include additional physical processes to account for the late evolution of satellites after infall (see Paper \RomanNumeralCaps{1}).
\end{enumerate}

In this work we also track the number densities of subhaloes that have reached the end of their dynamical time at each epoch. At the time a subhalo reaches its dynamical time the associated satellite galaxy `merges'\footnote{In a statistical model satellite galaxies are not strictly merging as there is no central galaxy to merge with, instead we collect statistics of merging satellites at each epoch. Using post-processing techniques we use these statistics obtain information on the average merging history of central galaxies.} with the central galaxy, during these mergers we inject a fraction (40\%) of the satellite mass to the intracluster medium \citep{Moster2018Emerge10}. 

\subsection{Satellite Quenching}
\label{subsec:SatelliteQuenching}

\RevHighlight{We improve the satellite quenching model from Paper \RomanNumeralCaps{1} in two regards. Firstly, we implement a pre-processed fraction inspired by the results of \cite{Wetzel2015SatelliteReionization} implementing a mass dependent fraction such that, a minimum of 30\% and a maximum of 60\% of galaxies are prepossessed with a transitional mass range between the extremes of $10^{6} M_{\odot}$ to $10^{8} M_{\odot}$. We also use the latest dynamical quenching results from \citet{Cowley2019TheSurvey} updating the model presented in Paper \RomanNumeralCaps{1} to include a redshift dependence in all quenching timescales, EQUATIONS, The parameters, $ \tau_q$, $\mathrm{t_q}$ and $\tau_f$ are from the \citet{Wetzel2013GalaxyUniverse} delayed-then-rapid-quenching model. The parameter $ \tau_q$ is the quenching timescale after infall, $\mathrm{t_q}$( = $\mathrm{t_{infall}} - \tau_q$) the look-back time at which a galaxy starts the fast quenching mode, and $\tau_f$ controls the fade time over which the galaxy quenches. The star formation rate of a satellite galaxy after infall is then given by, EQUATIONS.}{The satellite quenching model in \textsc{steel}, presented in Paper \RomanNumeralCaps{1}, has two components:
\begin{itemize}
    \item A delayed-then-rapid quenching model \citep{Wetzel2013GalaxyUniverse}, according to which satellite galaxies upon entering a halo continue to form stars as if they were on the star formation main sequence until their quenching timescale, $ \tau_q$, has elapsed. After a time $\tau_q$ the star formation rate of the satellites is rapidly quenched over the fading timescale $\tau_f$.\footnote{The absolute quenching time is given by $\mathrm{t_q}$( = $\mathrm{t_{infall}} - \tau_q$)}
    \item The second component used is the halo mass-dependant cutoff \citep{Fillingham2018EnvironmentalGalaxies}, which envisions that satellite galaxies below a given stellar mass (dependant on host halo mass) are immediately quenched. 
\end{itemize}
The delayed-then-rapid quenching model is improved using the latest dynamical quenching results from \citet{Cowley2019TheSurvey} updating the model presented in Paper \RomanNumeralCaps{1} to include a redshift dependence in all quenching timescales,
}

\begin{equation}
\begin{split}
\tau_{q,z} = \tau_q * (1+z_{infall})^{-3/2},\\
\tau_{f,z} = \tau_f * (1+z_{infall})^{-3/2}.
\end{split}
\end{equation}

\RevHighlight{}{Additionally, we include a pre-processed fraction, inspired by the results of \cite{Wetzel2015SatelliteReionization}, implementing a mass-dependent fraction such that a minimum of 30\% and a maximum of 60\% of galaxies are prepossessed with a transitional mass range between $10^{6} M_{\odot}$ to $10^{8} M_{\odot}$. The star formation rate of a satellite galaxy after infall is then given by}

\begin{equation}
\label{eqn:SFR_Quench}
SFR(t, t_{infall}, M_{*, infall}) = SFR_{t_{infall}}*
\begin{dcases}
1, & \text{} t > t_q \\
e^{\big[-\frac{t_q-t}{\tau_f}\big]}, & \text{} t < t_q.
\end{dcases}
\end{equation}

\subsection{Abundance Matching}
\label{subsec:AbundanceMatching}
In this work we populate dark matter haloes with galaxies using the abundance matching technique where galaxies are assigned to haloes by comparing the relative abundances of galaxies and haloes. For the abundance matching we use the central haloes from the halo mass function described in Section \ref{subsec:HMF}, and a subhalo mass function subdivided by redshift of infall generated from \textsc{steel}. Subhaloes are assumed to follow the central SMHM relation at infall. We simplify our abundance matching by using a frozen model such that baryonic evolution after infall (stripping, starformation, etc.) is not included. The latter assumption provides a good approximation as after infall the dominant factor determining the abundances of satellite galaxies is the dynamical time and not evolutionary processes (Paper \RomanNumeralCaps{1}).

To fit stellar mass functions over multiple epochs we convolve our halo mass functions with a parametric SMHM relation similar to that proposed by \cite{Moster2010},
\begin{equation}
\label{eqn:MosAbn}
\begin{split}
M_*(M_h, z) &= 2M_hN(z)\Big[ \Big( \frac{M_h}{M_{n}(z)}\Big) ^{- \beta(z)} + \Big( \frac{M_h}{M_{n}(z)}\Big)^{\gamma(z)} \Big ]^{-1}\\
N(z) &= N_{0.1} +N_z\Big(\frac{z-0.1}{z+1}\Big)\\
M_{n}(z) &= M_{n,0.1} +M_{n,z}\Big(\frac{z-0.1}{z+1}\Big)\\
\beta(z) &= \beta_{0.1} +\beta_z\Big(\frac{z-0.1}{z+1}\Big)\\
\gamma(z) &= \gamma_{0.1} +\gamma_z\Big(\frac{z-0.1}{z+1}\Big).
\end{split}
\end{equation}

In what follows we adopt both the cmodel and PyMorph SMF described in Section \ref{subsec:SMF} at redshift $z=0$ to constrain the parameters N, M, $\beta$, and $\gamma$ (normalization, knee, low mass slope, and high mass slope). We use only the central stellar mass function, using the \cite{Yang2012EvolutionHalos} central/satellite identification, and central halo mass function. The fit is performed using a Markov Chain Monte Carlo (MCMC), implemented using the \textsc{python} package \textsc{emcee} \citep{Foreman-Mackey2013Emcee:Hammer}, over a large parameter space ($P_{M, N, \beta, \gamma}$) covering all four parameters. Given a point in parameter space $P_{M_i, N_i, \beta_i, \gamma_i}$, the stellar mass function is constructed using the halo mass function and the SMHM relation. Each bin of parent halo mass is associated to a Gaussian distribution of stellar mass with scatter 0.15 dex. This distribution is multiplied by the halo mass number density to convert to galaxy number density which are added to the relevant stellar mass bins of the stellar mass function in construction. This operation is then repeated over all mass bins of the halo mass function to produce the complete central stellar mass function. For each point $P_{M_i, N_i, \beta_i, \gamma_i}$ in the parameter space, the stellar mass function associated to that point is compared via a likelyhood function to the observed stellar mass function to provide the MCMC with the probability that the given point is the `true' SMHM relationship. 

We then fit to the \cite{Davidzon2017TheSnapshots} data both uncorrected and corrected for the cmodel and PyMorph fits respectively (see Section \ref{subsec:SMF} for details). At high redshift we use the central and subhalo mass functions initializing satellites at infall as described above\footnote{Ideally, as for low redshift, we would use the centrals only as we are primarily concerned with the central SMHM relation. However, lacking a well-defined central stellar mass function at high redshift, this method represents a reliable way to extend the model to higher redshifts.}. For central haloes the method is the same as detailed above, however, as we use the total stellar mass functions at high redshift we also include the total unevolved surviving subhalo mass function in the abundance matching. 
\RevHighlight{We use the assumption that haloes are centrals at infall to initialize satellite galaxies.} 
{We assume that a halo before infall hosts a central galaxy; under this assumption we use the central SMHM relation to assign satellite galaxy stellar mass at the point of accretion.} For the latter we must have information about the redshift of infall for subhaloes. We obtain from \textsc{steel} the unevolved surviving subhalo mass function as contributed by each redshift of infall. Each contributing part is calculated using the SMHM relation at the redshift of infall and added to the central stellar mass function using the same method as with the centrals. The total stellar mass function is compared, at each redshift step available, to the data via the likely-hood function to give the probability that the given point is the `true' evolution parameters. The abundance matching best-fit parameters and associated errors for both the cmodel and PyMorph are given in Table \ref{tab:AbnResult}, and plots showing the cross sections of the parameter space are shown in Appendix \ref{app:ABN_MCMC}.

\begin{table*}
\centering
\begin{tabular}{l|llllllll}
        & M\_n & N     & $\beta$ & $\gamma$ & $M_{n,z}$ & N\_z   & $\beta_z$ & $\gamma_z$ \\ \hline
\\
cmodel  & $11.91_{-0.34}^{+0.40}$ & $0.029_{-0.013}^{+0.018}$ & $2.09_{-1.02}^{+1.21}$    & $0.64_{-0.10}^{+0.11}$     & $0.52_{-0.19}^{+0.24}$       & $-0.018_{-0.004}^{+0.005}$ & $-1.03_{-0.34}^{+0.049}$     & $0.084_{-0.14}^{+0.20}$      \\
\\
PyMorph & $11.92_{-0.36}^{+0.39}$ & $0.032_{-0.012}^{+0.016}$ & $1.64_{-0.73}^{+0.85}$     & $0.53_{-0.11}^{+0.11}$     & $0.58_{0.19}^{+0.15}$        & $-0.014_{-0.006}^{+0.007}$ & $-0.69_{-0.36}^{+0.29}$      & $0.03_{-0.147}^{+0.154}$      
\end{tabular}
\caption{The abundance matching results for the cmodel and PyMorph data. The errors are the 16th and 86th percentile from the MCMC fiting.}
\label{tab:AbnResult}
\end{table*}

\begin{figure*}
	\centering
	\includegraphics[width = \linewidth]{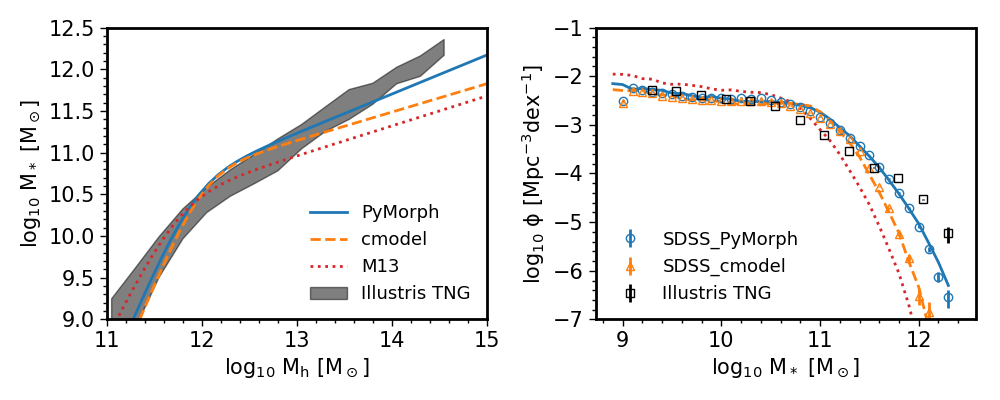}
    \caption{Left: The SMHM relation at redshift $z=0.1$. The PyMorph (blue solid line) and cmodel (orange dashed line) fits from this work are both for central haloes/galaxies, the fit from \citet{Moster2013} (hereafter M13, red dotted line) is for all haloes/galaxies. The grey band is the relation from Illustris TNG100. Right: Stellar mass functions created using the central halo mass function and the three SMHM relations compared to PyMorph (blue circles) and cmodel (orange triangles) central stellar mass functions. The black squares are the stellar mass function from Illustris TNG100.}
	\label{fig:AbnMtch_lz}
\end{figure*}

In Figure \ref{fig:AbnMtch_lz} we show the results of our abundance matching to the PyMorph and cmodel central stellar mass functions. The PyMorph fit is steeper above the knee compared to either the cmodel or the \citet{Moster2013} model fits, as expected given the larger number density of massive galaxies found applying the S\'ersic-Exponential model \citep[eg.,][]{Shankar2014, Kravtsov2018StellarHalos}. The low mass slope for both PyMorph and cmodel are almost identical as the galaxies in this range are not affected by the photometric choice. Differences between the fits from this work and \citet{Moster2013} are due to our selection of using only central haloes/galaxies as opposed to the total population, and the stellar mass functions shown in the right-hand panel are lower than even cmodel are are therefore missing massive galaxies.

\section{Results}
\label{sec:Results}

\subsection{Halo Merger Rates}
\label{subsec:HaloMergerRate}

\begin{figure}
	\centering
	\includegraphics[width = \linewidth]{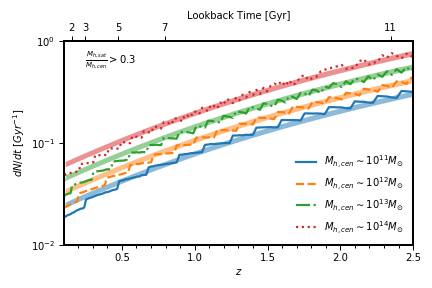}
    \caption{The evolution of merger rate per Gyr at fixed halo mass. Lines are from \textsc{steel}, shaded bands are the analytic fits from \citet{Fakhouri2010TheSimulations}. Halo masses shown are $M_{h,cen}: 10^{11}, 10^{12}, 10^{13}, 10^{14} M_{\odot} h^{-1}$ as labelled.}
	\label{fig:HaloMergerRate}
\end{figure}

\textsc{steel} implements a statistical dark matter merging history, thus as a very first step we check STEEL's performance on reproducing halo merger rates as extracted from N-body dark matter-only simulations. We explore the evolution of the merger rate of haloes with a mass ratio greater than $f = M_{h, sat} / M_{h, cen}$. The merger rate is calculated from \textsc{steel} by integrating the ``unevolved sub-halo mass function accretion'' ($\delta$USHMF) above the mass ratio limit, 
\begin{equation}
\begin{split}
\frac{dN}{dz}\Big(M_{h, cen}\Big) =& \\
\int^{\infty}_{M_{h,cen}*f} \delta &USHMF([z, z+\delta z], M_{h,cen}, M_{h, sat}) dM_{h,sat}.
\end{split}
\end{equation}
In Figure \ref{fig:HaloMergerRate} the merger rate from \textsc{steel}, shown by lines, is in good agreement with the best-fit merger rate relations from the Millennium simulation given by \citet{Fakhouri2010TheSimulations}, shown as shaded regions\footnote{It should be taken into consideration that the results from STEEL presented here and the fits from \citet{Fakhouri2010TheSimulations} are in different cosmologies. We show \textsc{steel} halo accretion using the Millennium cosmology used in the aforementioned work in Appendix \ref{app:CosmoAlt}.}. The slight deviation at low redshift derives from the predicted growth history of our input potential wells given by \cite{vandenBosch2014ComingWells}, a lower mass growth rate leads to a lower accretion rate. \RevHighlight{}{This deviation is due to differences in the algorithms used to link haloes between simulation outputs and build merger trees used by \citet{Fakhouri2010TheSimulations} and \cite{vandenBosch2014ComingWells}}

\subsection{High Redshift Clusters}
\label{subsec:HighRedshiftClusters}
We here extend the group and cluster satellite richness analysis from Paper \RomanNumeralCaps{1} to high redshift. In Paper \RomanNumeralCaps{1} it was found that dynamical friction and, to a second order, abundance matching, are the dominant factors in the distribution of satellite galaxies in groups and clusters above $M_{*,sat} > 10^{10} M_{\odot}$. In this section, for a more rounded view of the satellite galaxy population, we display the results for the full \textsc{steel} model which includes star formation, dynamical quenching and stripping to evolve satellites after infall. The latter effects, despite being of lower order than dynamical friction or abundance matching, are included to be able to compare to data other than cluster richness, such as the satellite specific star formation rate distribution.

\begin{figure*}
	\centering
	\includegraphics[width = \linewidth]{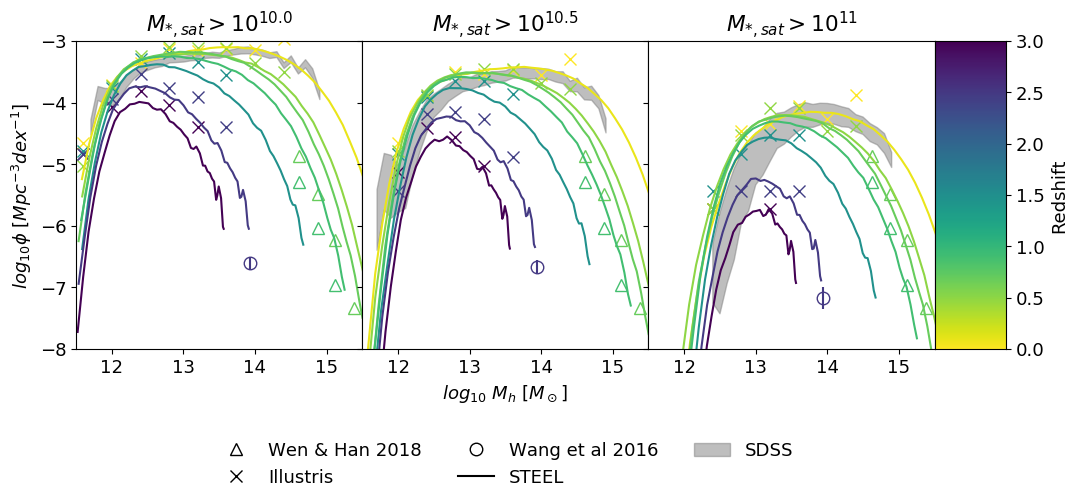}
    \caption{The number-density distribution of satellites per parent halo mass predicted from \textsc{steel}, using the PyMorph SMHM relation, at multiple redshift epochs (solid lines). The grey band is the data from SDSS at redshift $z=0.1$. Also included are the high redshift cluster data from \citet{Wang2016DiscoveryZ=2.506} (circles) and \citet{Wen2018ARedshifts} (triangles). We also compare to the outputs from the Illustris simulation using the TNG100 data (crosses). Each data point and line are given a colour associated to their redshift (the bar on the right provides the color coding key).}
	\label{fig:HighZClusters}
\end{figure*}

\RevHighlight{Figure \ref{fig:HighZClusters} shows the satellite number density per halo mass bin for three satellite mass cuts.}{Figure \ref{fig:HighZClusters} shows the satellite number density per halo mass bin. For each central halo mass the cosmic number density, similarly to the number density presented in the cumulative stellar mass functions, is calculated for satellites above a mass threshold for each halo mass bin.}
The predicted halo richness from \textsc{steel}, using the PyMorph SMHM relation, is shown in this plot as solid lines. Low redshift SDSS data are shown as a grey band, cluster data detailed in Section \ref{subsec:Clusters} are open symbols. The predictions from the Illustris TNG100 simulation \citep{Nelson2018FirstBimodality, Springel2018FirstClustering} are shown with crosses. The markers and lines in the figure are colour coded based on redshift, as indicated by the colour bar on the right.

\RevHighlight{}{In Figure \ref{fig:HighZClusters} we see that at higher redshift there are fewer massive satellites overall. This reduction is caused by several contributing effects. Firstly, at high redshift there is an absence of high mass haloes, which have not had time to form. Due to the lack of hosts at higher masses, the right hand side of the distribution tightens. Secondly, at high redshift the halo mass function is lower at any given mass, causing a shift of the satellite host halo distributions towards lower masses. Finally, as the process of formation and merging takes several gigayears to complete massive satellites are found less frequently at high redshift, and thus the number densities of the most massive satellites reduce faster than the lower mass ones.}

We show \textsc{steel} is a good match to the \cite{Wang2016DiscoveryZ=2.506} Cluster at redshift $z = 2.5$. We also achieve an adequate match to the cluster survey from \cite{Wen2018ARedshifts}, especially in the mass range above $\mathrm{M_{sun}} > 10^{10.5} M_{\odot}$. We also achieve similar results to the Illustris TNG100 simulation, though the TNG100 output is marginally higher at all redshifts. \textsc{steel} improves upon TNG100 in terms of the shape and breadth of this distribution. For example, at high mass and redshift \textsc{steel} resolves the turnover for satellites above $\mathrm{M_{sun}} >10^{11} M_{\odot}$, whereas TNG100 is too limited in volume to cover the high mass ranges. The limitations in volume prevent TNG100 from simulating clusters such as those presented in Figure \ref{fig:HighZClusters} \citep{Wang2016DiscoveryZ=2.506, Wen2018ARedshifts}. In this respect \textsc{steel} becomes an excellent bridge between the capabilities of a high resolution hydrodynamical simulation and the massive cluster observations at high redshift.

\subsection{Connecting Central Mass Accretion to Star Formation Rate }
\label{subsec:CentMassAcc}

\begin{figure*}
	\centering
	\includegraphics[width = \linewidth]{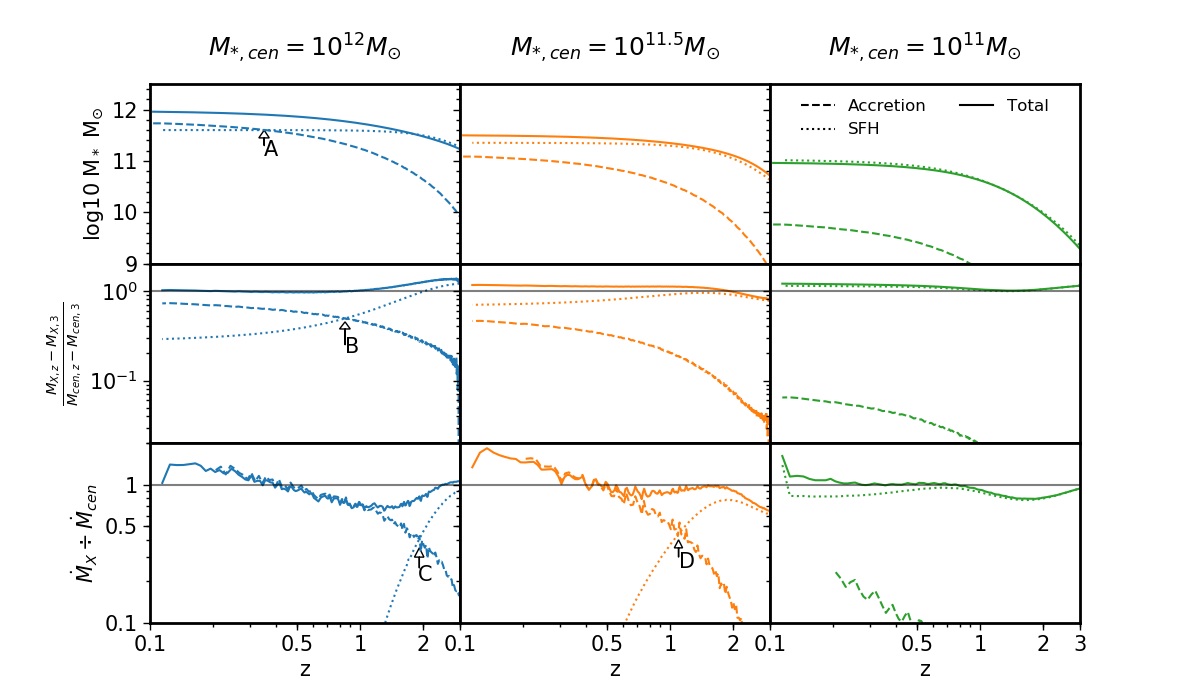}
    \caption{Three `mass tracks' are shown that have central galaxy masses at redshift $z = 0.1$ of $M_{*,cen}$ = $10^{12}$, $10^{11.5}$, and $10^{11}$ $[M_{\odot}]$ in blue orange and green respectively. The satellite galaxy accretion is shown for evolved satellites with a dashed line, and the mass from star formation shown with a dotted line. The top panels show the total mass of the central (solid lines) and the total mass gained from accretion or star formation. The middle panels show the fraction of the total galaxy mass formed from satellite accretion or star formation since redshift $z=3$. The bottom panels show the ratio of the mass accretion rate from satellite galaxies, the star formation rate, and the mass growth rate of the central galaxy predicted by abundance matching. The black horizontal lines in the second and third rows are at unity. The solid lines showing the sum of the other two factors should be close to or on the unity lines. The labels A \& B point to where the cumulative mass from accretion overtakes the cumulative mass from star formation. The labels C \& D point to where the instantaneous accretion overtakes the star formation rate.}
	\label{fig:SatelliteAccretion}
\end{figure*}

We calculate from \textsc{steel}, using the PyMorph SMHM relation, the relative contributions to the average stellar mass growth of central galaxies from satellites and star formation history, as shown in Figure \ref{fig:SatelliteAccretion} . For three galaxy mass bins ($10^{11}$,$10^{11.5}$,$10^{12}$ $M_{\odot}$) selected at $z = 0.1$, the average growth history (total, solid lines) is derived by following the host halo-mass track, and the stellar-mass track is implied by imposing abundance matching at all redshifts. The stellar mass history assigned by abundance matching, is naturally independent of any galaxy merger modelling assumptions from \textsc{steel}. The total accretion from satellites (accretion, dashed lines) is computed from the expected satellite accretion along halo mass tracks. For each galaxy a star formation history (SFH, dotted lines) \RevHighlight{is calculated initializing the galaxies mass at redshift $z = 5$ using abundance matching, the method for deriving the star formation rate-stellar mass relation used in described below.}
{may then be calculated. The star formation rate is tuned such that it provides the correct star formation history to account for the difference between the mass growth expected from abundance matching and the cumulative satellite stellar mass accretion}\footnote{This method directly links the star formation rate to the accreted mass from satellites. However, in our model satellites to grow in mass after infall (i.e., `non-frozen' to follow the terminology of Paper \RomanNumeralCaps{1}), we therefore recalculate the full satellite accretion onto the central galaxies updating their mass using the new star formation rate. Using the updated accretion the star formation rate is recalculated beginning an iterative process. However, this iterative process of recalculation ends after one loop as the re-derived accretion is found to be nearly identical, as expected from the results of Paper \RomanNumeralCaps{1}.}. When calculating this difference we also take into account the stellar mass loss rate (MLR) due to stellar recycling using the relations from \citet{Moster2018Emerge10},

\begin{equation}
\label{eqn:f_ml}
f(\tau_{ml}) = 0.05 \ln \Big(\frac{\tau_{ml}}{1.4 Myr}+1\Big) ,
\end{equation}

\begin{equation}
\label{eqn:MLR}
MLR(t) = \frac{ \sum_{t' = t_{infall}}^{t} SFH(t')(f[t' - (t-\delta t)]-f[t' - t]) }{\delta t} .
\end{equation}

The star formation rate - stellar mass relation derived from this method is fitted with a double power law that evolves with redshift (for more details on the fit see Appendix \ref{app:SFR}), as shown in Figure \ref{fig:SFR_DPL}. At lower redshift the normalization decreases, the peak of the distribution shifts to lower masses, and the turnover after the peak is steeper. In Figure \ref{fig:SFR_DPL} we also show the same three galaxy population tracks from Figure \ref{fig:SatelliteAccretion}, discussed below, as black lines. These tracks show how the galaxy population evolves in SFR with redshift. The population tracks show a gradual increase in SFR and then a turnover before dropping sharply, as they transition to a satellite accretion-dominated regime. It is found that smaller galaxies grow for longer timescales with increasing star formation, whilst larger galaxies start with higher star formation rate and transition to an accretion-dominated phase much earlier in time. 

The top row of Figure \ref{fig:SatelliteAccretion} shows the total mass of the galaxy and the total contributed by both accretion and the star-formation history. The middle row shows the fractional contributions from star-formation and satellite accretion from $z = 3$. The bottom row shows the instantaneous mass growth from star formation and satellite accretion. There are two definitions we can use to determine the epoch after which a galaxy transitions into a merger-dominated state. Firstly, we could define the ``cumulative transition'' as when the galaxy has accreted more mass than it has created from star-formation processes (Points A \& B). Secondly, we define the ``instantaneous transition'' as  the epoch when the growth rate from mergers overtakes the growth rate from star-formation (Points C \& D). More massive galaxies transition earlier to merger dominated growth under both definitions. However, all galaxies transition earlier under the second (instantaneous) definition. The masses shown in Figure \ref{fig:SatelliteAccretion} show three cases of relevant galaxy accretion tracks. The $M^{z=0}_* = 10^{12} M_{\odot}$ galaxy growth curve at low redshift is always dominated by satellite accretion. In the top and middle rows we see that more mass has been accreted than produced by star formation, and in the bottom row we see the accretion rate overtook the star formation rate at redshift $z=2$. The $M^{z=0}_* = 10^{11.5} M_{\odot}$ galaxy growth curve has more mass created from star formation than satellite accretion. However, the galaxy population has a higher rate of accretion rate than star formation rate since redshift $z = 1$. The final population shown at $M^{z=0}_* = 10^{11} M_{\odot}$ is star formation dominated under both cumulative and instantaneous definitions. At redshift $z = 0$ we find the transition masses for the total mass ratio and the instantaneous ratio to be at $M_* = 10^{11.7} M_{\odot}$ and $M_* = 10^{11.1} M_{\odot}$ respectively.

\begin{figure}
	\centering
	\includegraphics[width = \linewidth]{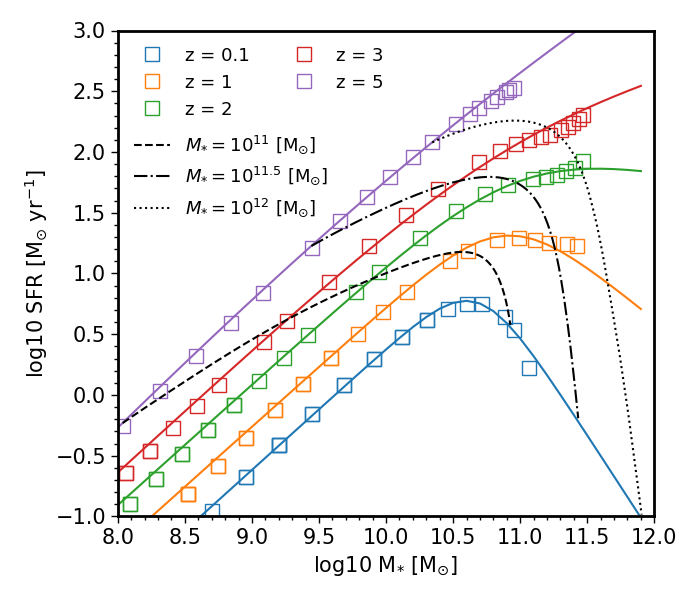}
    \caption{The star formation rate - stellar mass relation derived from following central galaxy populations along halo mass histories at redshifts $z = 0.1, 1, 2, 3, 5$. The data extracted from the post-processing of \textsc{steel} are shown by coloured crosses and the double power-law fits are shown as lines in corresponding colours. The three black lines are the evolution of the galaxy populations selected at redshift $z=0.1$ with masses $M_* = 10^{11}, 10^{11.5}, 10^{12} [M_{\odot}]$  presented in Figure \ref{fig:SatelliteAccretion}.}
	\label{fig:SFR_DPL}
\end{figure}

\begin{figure*}
	\centering
	\includegraphics[width = \linewidth]{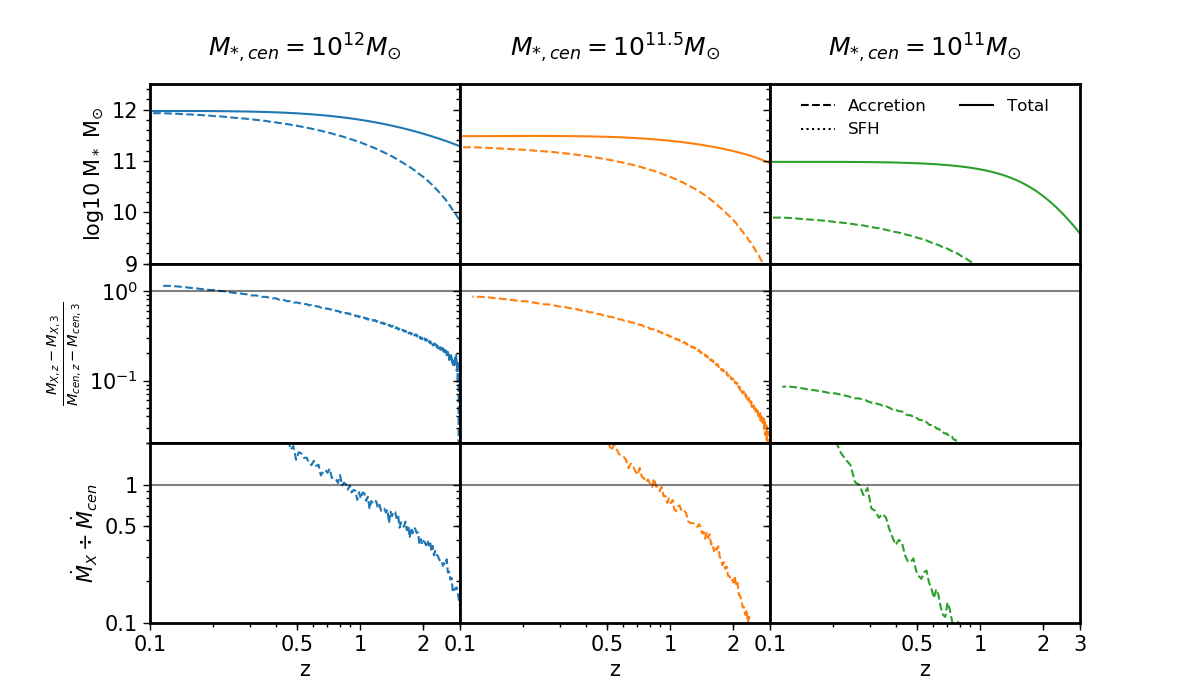}
    \caption{Same format as Figure \ref{fig:SatelliteAccretion} but for the cmodel photometry. It is clear that this model is internally nonphysical as the accretion via satellites (dashed lines) rapidly overshoots the total growth in stellar mass (solid lines) implied by the underlying growth host halo growth, as evident in the middle and bottom rows.}
	\label{fig:SatelliteAccretioncMod}
\end{figure*}

We show in Figure \ref{fig:SatelliteAccretioncMod} the satellite accretion for the cmodel abundance matching using the same template as Figure \ref{fig:SatelliteAccretion}. In Figure  \ref{fig:SatelliteAccretioncMod} we obtain a lower limit for the accretion rate by including stripping but not star-formation in the satellites thus minimizing their mass through environmental processes. We find for the high mass galaxies, which are above the knee of the SMHM relation, even the lower limit for the accretion has an instantaneous rate greater than the growth rate of the galaxy as seen in the bottom row. This makes the cmodel SMHM relation used within our dark matter accretion model \textit{non-physical}: steeper SMHM relations, such as the one found with the PyMorph photometry, are favoured by hierarchical assembly. We recall, as explained in Figure \ref{fig:Simple_Cartoon}, that too flat SMHM relations introduce global stellar mass growth histories that are even lower than what is expected from total satellite accretion rendering the models internally inconsistent. \RevHighlight{}{For completeness we also tested a range of dynamical time options. In all cases, even when the merging time is increased by a factor of two}\footnote{Dynamical time factors higher than 2 are shown to not reproduce the satellite distributions in Paper \RomanNumeralCaps{1}}\RevHighlight{}{, the accretion rate exceeds the growth rate and cmodel photometries can be excluded.  Further to this we also tested variations on the mass loss in mergers and the amount of mass lost to tidal stripping of the satellites; in the case where the tidal is doubled with respect to the reference model and the mass loss during a merger is set to 60\% up from 40\%, the cmodel photometries remain internally inconsistent. We are confident that under all circumstances cmodel photometries can be considered internally inconsistent.}

\subsection{Specific Star Formation Rate Distribution}
\label{subsec:sSFR}

\begin{figure*}
	\centering
	\includegraphics[width = \linewidth]{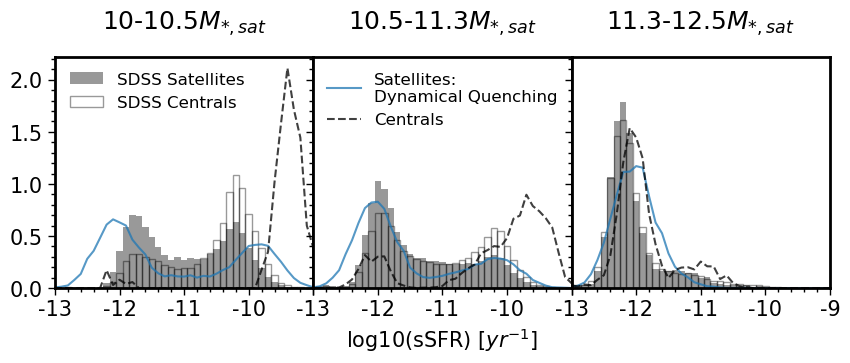}
    \caption{We show the sSFR of satellites and centrals compared to SDSS in three mass bins selected to mirror proposed breaks in the galaxy main sequence. The SDSS data for satellites and centrals are filled and unfilled histograms respectively. The \textsc{steel} result for the satellites is the solid blue line and the post processed central result is the dashed black line.}
	\label{fig:sSFR}
\end{figure*}

Figure \ref{fig:sSFR} shows the specific star formation rate distribution of satellites in three mass ranges, as labelled, chosen to probe transitions found in observational data \citep{Bernardi2011EvidenceRelations, Bernardi2014SystematicMorphology, Cappellari2013TheFunction}. The solid blue line and the dashed black lines show the satellite and central sSFR from \textsc{steel}, respectively, while the grey histogram shows the satellites from SDSS and the unfilled histogram shows the centrals in SDSS. 

\textsc{steel} accurately captures the key trends in the distributions, such as bimodality, which is seen in both the central and satellite populations. The central population below $M_{*} = 10^{10.5}$ [M$_{\odot}$] is mostly star-forming whereas the satellites show signs of quenching. In the intermediate-mass range a fraction of the centrals become quenched and the satellites show a strong quenching effect. In the highest mass range all galaxies show strong quenching features with little star-formation. Whilst still not an exact match to the SDSS distribution, we find that including a redshift dependence in the dynamical quenching provides a better fit than the model used in Paper \RomanNumeralCaps{1}. The central sSFR is calculated using the star formation rate presented in Section \ref{subsec:CentMassAcc}, which use the PyMorph SMHM relation. Each central mass is assigned a star formation rate with a scatter of 0.2 dex. To account for the fraction of galaxies that are quenched via mergers at each stellar mass we modify the assigned star formation rates by setting a fraction of galaxies equal to the elliptical fraction from Section \ref{subsec:Morph} to have a sSFR of $10^{-12}$ [$yr^{-1}$] with a scatter of 0.2 dex and in turn increase the star formation rate of the remaining galaxies to maintain the same average star formation rate for the population. This approach tests if mergers alone can account for the bimodality found in the central sSFR, the high mass centrals $ > 10^{11.3}$ [M$_{\odot}$], but produces an inadequate fit to the SDSS centrals at masses lower than $10^{10.5}$ [M$_{\odot}$]. The discrepancies in the location of the star-forming population are likely caused by the imperfect fit to observed SFR as seen in \ref{fig:SFR_L18} and the deficit of quenched galaxies in the lower mass cuts is likely due to causes of quenching that are not merger related (e.g., AGN feedback).

\section{Discussion}
\label{sec:Discussion}

Using \textsc{steel} we have presented a consistent picture of group and cluster richness across several orders of magnitude in mass, and satellite accretion histories over 11 Gyr of the Universe's history. It is essential for a model that aims to predict the hierarchical growth of galaxies, that both the central and satellite stellar mass functions are well reproduced at all redshifts. \textsc{steel} uses state-of-the-art statistical accretion histories and powerful abundance matching techniques to ensure we have the essential consistency with observed galaxy number densities by design. 

\subsection{High Redshift Clusters}

Galaxy groups and clusters represent an excellent laboratory to test theories on galaxy evolution. The rich cluster environments are observable up to high redshift and contain some of the most massive galaxies. Exploring the richness of the environments around massive galaxies provides an excellent constraint to  hierarchical assembly predicted by $\Lambda$CDM cosmology at the most extreme masses \citep{Shankar2015}. In this work we show a singular cluster reported in \citet{Wang2016DiscoveryZ=2.506}. Other models \cite[e.g.][]{Henriques2015GalaxyMasses} have been unable to reconcile this cluster within a $\Lambda$CDM framework. We found in Figure \ref{fig:HighZClusters} that \textsc{steel} is able to predict the existance of such massive objects. However we concur with \citet{Wang2016DiscoveryZ=2.506} that these objects are rare and their absence in traditional simulations could be simply attributed to poor statistics and not necessarily to the implied physical model. With large-scale surveys such as EUCLID coming online, a well-tuned statistical model could more easily place robust constraints on high-redshift cluster formation. 

\subsection{Central Assembly}

In Section \ref{subsec:CentMassAcc} we found one of the major factors in regulating the in-situ and ex-situ accretion pathways to be the \textit{shape} of the SMHM relation. A shallower low-mass slope causes larger amounts of satellite accretion as smaller haloes, with much higher number density, are initialized with larger satellite galaxies. Similarly to \citet{Shankar2006NewFormation} \& \citet{Moster2018Emerge10}, we find the high mass slope to undergo only a small amount of evolution with increasing redshift, this implies the growth of central galaxies is directly linked to the steepness of the high mass slope and the growth of the host halo. The flatter the high mass slope of the SMHM relation, the less growth is expected in stellar mass following the assembly of the host dark matter halo. In turn, a weak evolution in the stellar mass content of the central galaxy can be in tension with what is expected from satellite accretion, especially for the most massive galaxies. We discussed that the slope of the high-mass end of the stellar mass function and implied slope of the SMHM relation strongly depend on the choice of light profile, background subtraction, and mass-to-light ratios. However, not all resulting stellar mass functions provide physically self-consistent results in a LCDM Universe. Steeper SMHM relations, such as those predicted by PyMorph-based stellar mass functions \citep{Bernardi2013TheProfile}, produce more consistent central and satellite accretion stellar mass growths. In addition to models with different SMHM slopes, we also tested models with the dynamical time varied by $\pm$20$\%$, within the range of possible dynamical times predicted in Paper \RomanNumeralCaps{1} constrained by satellite richness. This relatively modest alteration has a minor effect on the satellite accretion rate and mass contribution to the central. In this work we find the transitional stellar mass, above which dry mergers progressively become the major contributor to galaxy growth, to be $M_{*} = 10^{11.1}$, see Figure \ref{fig:SatelliteAccretion}. The latter is consistent with previous findings \citep[e.g.,][]{Bernardi2011EvidenceRelations, Cappellari2013EffectEvolution, Shankar2013SizeUniverse}.

\begin{figure}
	\centering
	\includegraphics[width = \linewidth]{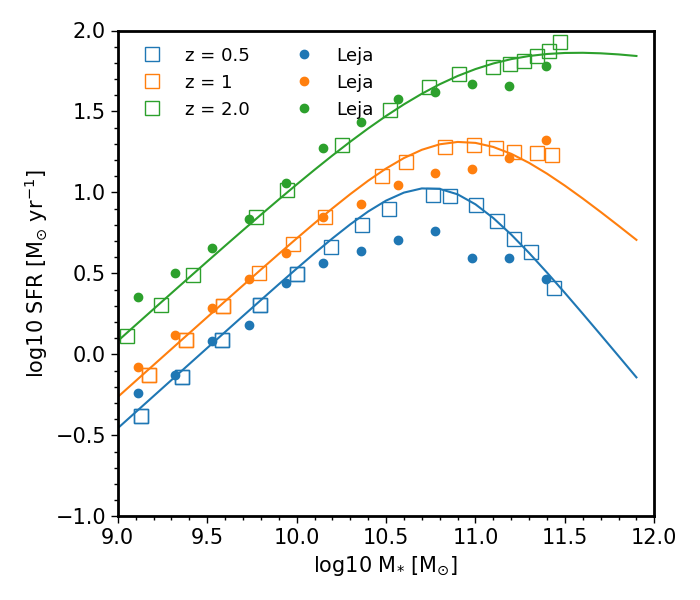}
    \caption{We show the star formation rate - stellar mass relationship from Figure \ref{fig:SFR_DPL} at redshifts z = 0.5, 1, 2 (blue orange and green respectively, \textsc{steel} data are crosses and fits are solid lines). In this plot we compare with the observed star formation rate from \citet{Leja2018ANSURVEY} shown as filled circles with corresponding colours denoting corresponding redshift.}
	\label{fig:SFR_L18}
\end{figure}

By following the statistical dark matter accretion histories we were able to use the central mass tracks and abundance matching to obtain a growth history for central galaxies. Subtracting from the latter at each time step the cumulative stellar mass from satellite accretion, we created a `star formation rate' interpreted as the remaining mass required to build the central mass. Our methodology is similar to the continuity approach based on \citet{Leja2015ReconcilingFunction} used in Paper \RomanNumeralCaps{1}, but with the key difference that here we follow halo growth instead of galaxy number density. 
\RevHighlight{The resulting star formation rate for galaxies below the turnover is higher than the star formation rate from \citet{Tomczak2014GALAXY}, used in Paper \RomanNumeralCaps{1}, at high redshift by 0.1-0.2 dex and lower at low redshift $z < 1$ by 0.3 dex and also shows a very different turnover mass at all redshifts.}
{The resulting star formation rate for galaxies is notably different to that of \citet{Tomczak2014GALAXY}, used in Paper \RomanNumeralCaps{1}. At all redshifts the turnover is notably different, with SFR for masses above the turnover decreasing sharply at low redshift. For masses below the turnover, at $z < 1$ the SFR is lower by 0.3 dex, and at $z > 1$ the SFR is higher by 0.1-0.2 dex.}
Recent work, where the star formation histories are properly accounted for when measuring star formation rates, has suggested that the previous determinations of star formation rates using UV+IR are 0.1 to 1 dex too high \citep{Leja2018ANSURVEY} and cannot be reconciled with the growth of the stellar mass function \citep{Leja2015ReconcilingFunction, Lapi2017StellarEquation}. Our star formation rate is consistent with the results of \citet{Leja2018ANSURVEY}, as reported in Figure \ref{fig:SFR_L18}. The excellent match to Leja et. al.'s independent estimates further supports the idea that a more robust method to derive more reliable star formation rates is to follow galaxy assembly along host halo growth histories \citep[see e.g.,][]{Moster2018Emerge10}.

\subsection{Central Morphologies}
\label{subsec:Morph}
\begin{figure}
	\centering
	\includegraphics[width = \linewidth]{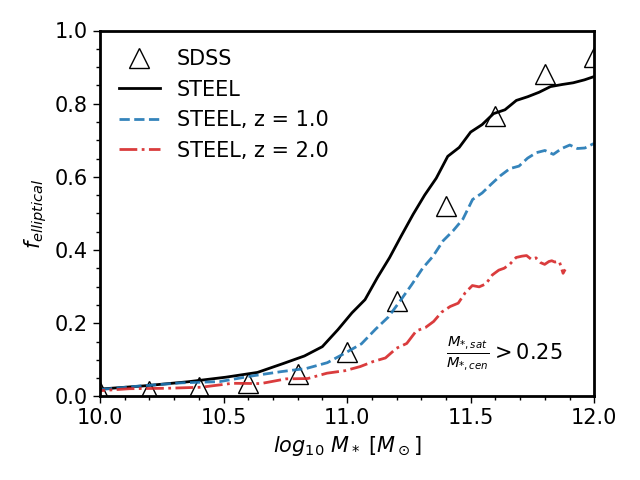}
    \caption{We show at three redshift steps the predicted fraction of ellipticals as a function of stellar mass. The lines are the predictions from \textsc{steel} and the tringles are the T-Type selected elliptical fraction from SDSS at redshift $z = 0.1$.}
	\label{fig:Morphologies}
\end{figure}

Mergers are thought to be one of the drivers for morphological transformation, size growth and other galaxy changes \citep{Bournaud2007, Hopkins2009TheDemographics, Hopkins2010MergersFunctions, Shankar2011SizeUniverse, Fontanot2015OnMergers}. Broadly inspired by the results of hydrodynamic simulations, a number of analytically-based cosmological models have generally assumed that major mergers, in particular, with a mass ratio of at least $M_{sat}/M_{cen} > 0.25$, are effective in destroying disks and in forming ellipticals \citep{Baugh2006AApproach, Malbon2007BlackFormation, Bower2010TheFormation}. Given the very promising results of \textsc{steel} in predicting satellite number densties in different environments and epochs, we here take a step further and explore whether \textsc{steel}'s cumulative number of major mergers is able to account for the local fraction of elliptical galaxies. For each central mass track we evolve the fraction of galaxies that have had a merger with stellar mass ratio greater than 0.25 since redshift $z=3$. Figure \ref{fig:Morphologies} shows the probability/fraction of central galaxies that have experienced a merger above the mass threshold of 0.3 at redshifts $z = 0.1, 0.65, 1.75$, while the black triangles show the T-Type-selected elliptical fraction from the SDSS catalogue. We find that applying this simple recipe to the merging number densities from \textsc{steel} creates a good match to the elliptical fraction in the local universe.  

Despite the noticeably good agreement between model predictions and data in Figure \ref{fig:Morphologies}, we stress that different input SMHM relations can, as proven in this work, substantially affect the accretion rate which in turn will modify the number of galaxies experiencing major mergers. It follows that any cosmological galaxy evolution model that uses mergers as a physical driver for galaxy transformation should first simultaneously and self-consistently closely reproduce stellar mass functions, the SMHM relation, and satellite distributions at high redshift.

\section{Conclusions}
\label{sec:Conclusions}

In this second paper on our STastical sEmi-Empirical modeL, \textsc{steel}, we proved that \textsc{steel} can successfully reproduce galaxy satellite richness also at high redshifts. Its innovative design, unconstrained by volume or mass resolution, allows \textsc{steel} to predict the number densties of even the rarest objects in the Universe at the highest redshifts, a fundamental test for dark matter and galaxy evolution theories though currently inaccessible by cosmological hydrodynamic simulations.

Given the success of \textsc{steel} in reproducing satellite richness at different cosmic epochs and environments, we can in turn predict more reliable galaxy merger rates, using central growth rates implied by the central mass track of our statistical dark matter accretion histories and abundance matching. We found that SMHM relations with shallow high-mass slopes create central growth histories that are physically inconsistent with the expected satellite merger rate. We found that steeper SMHM relations at the high mass end, as induced by the latest determinations of the stellar mass functions based on Sersic-Exponential photometry, are favoured against shallower SMHM relations, based on outdated determinations of the stellar mass function. The total stellar mass growth of a galaxy is mostly due to satellite mergers and/or star formation. A flatter SMHM relation, however, naturally implies, for a given increase in host halo mass, a much weaker growth in the stellar mass of the central galaxy than in the case of a steep SMHM relation. The accretion via satellites could then be substantial enough to overshoot the moderate growth in the central galaxy rendering the model internally physically inconsistent.

By safely assuming the difference in central growth rate ($\dot{\mathrm{M}})$ and satellite accretion rate is attributable to the star formation in the central galaxy, we predict star formation histories and a star formation rate-stellar mass relations. The latter approach is qualitatively similar to a continuity equation \citep[e.g.,][and Paper \RomanNumeralCaps{1}]{Leja2018ANSURVEY}, but more accurate as it is developed along the accretion tracks of host haloes so better follows galaxy populations. We find our resulting star formation rates to be in excellent agreement with the latest cutting-edge observational measurements by \citet{Leja2018ANSURVEY}, based on multi-parameter Bayesian analysis.

Finally, following traditional models of galaxy evolution, we use our improved galaxy merger rate to predict the fraction of central galaxies as a function of mass that have been transformed into ellipticals via major mergers, (where the stellar mass ratio of the central to satellite is greater than 1/3). We find this fraction to be in excellent agreement with centrals selected as ellipticals via T-Type in SDSS. We use this elliptical fraction and our derived star-formation rate to create a distribution of specific star formation rates. We find this basic and common assumption to form ellipticals in analytic cosmological models to be sufficient to also reproduce the bi-modality in star formation rates of massive galaxies above $M_* > 10^{11.3}$ $M_{\odot}$. Below this stellar mass threshold, we find a too high fraction of star forming galaxies, which implies additional quenching mechanisms, beside major mergers, must be included in the models. 

Our results are of the utmost importance to predict robust and self-consistent SMHM relations and to generate reliable mock catalogues for the next generation of extra-galactic surveys such as Euclid and LSST.

\section*{Acknowledgements}

We acknowledge extensive use of the Python libraries astropy, matplotlib, numpy, pandas, and scipy. PJG acknowledges support from the STFC for funding this PhD. FS acknowledges partial support from a Leverhume Trust Research Fellowship \& from the European Union's Horizon 2020 programme under the AHEAD project (grant agreement \#654215).




\bibliographystyle{mnras}
\bibliography{Mendeley} 



\appendix

\section{Abundance Matching MCMC}
\label{app:ABN_MCMC}

\begin{figure*}
	\centering
	\includegraphics[width = \linewidth]{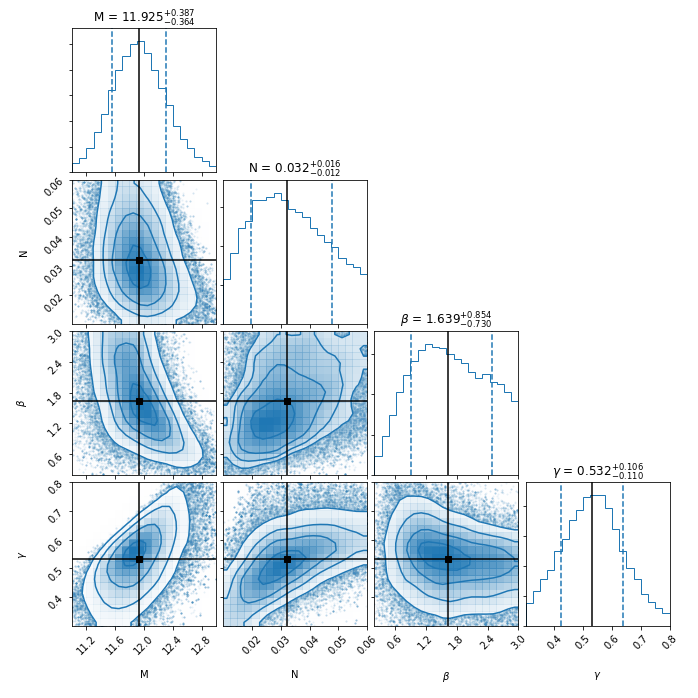}
    \caption{We show the MCMC parameter space for the redshift $z = 0.1$ fit. The position of the knee (M), the normalization (N) the low mass slope ($\beta$) and the high mass slope ($\gamma$) are shown from left to right. Columns are titled with the best fit values and 16th/84th percentile errors. The black lines show the best fit value with a black square at intersections, the 16th/84th percentiles are shown with blue dashed lines on the histograms.}
	\label{fig:MCMC_lz}
\end{figure*}

Figure \ref{fig:MCMC_lz} shows the redshift $z = 0.1$ output from the MCMC abundance matching fits. It becomes immediately obvious that the low mass slope ($\beta$) is poorly constrained however the impact on the SMF is limited within the margin of error. The position of the knee (M) is well constrained against both the normalization (N) and the high mass slope ($\gamma$). The shape of the constraint between the normalization and gamma emanates from the need to produce high mass galaxies, if the normalization is decreased the slope must increase to ensure enough haloes produce massive galaxies.

\begin{figure*}
	\centering
	\includegraphics[width = \linewidth]{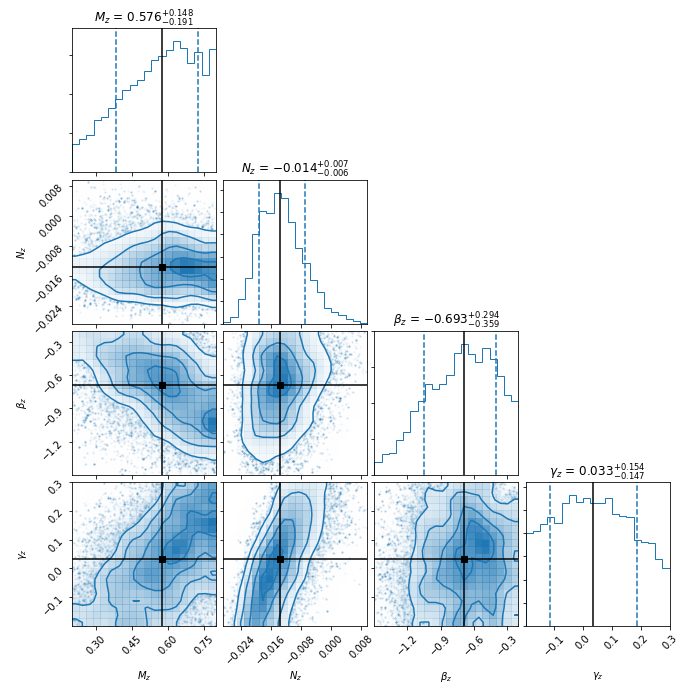}
    \caption{We show the MCMC parameter space for the high redshift $z > 0.1$ fit. The evolution of: the position of the knee ($M_z$), the normalization ($N_z$) the low mass slope ($\beta_z$) and the high mass slope ($\gamma_z$) are shown from left to right. Columns are titled with the best fit values and 16th/84th percentile errors. The black lines show the best fit value with a black square at intersections, the 16th/84th percentiles are shown with blue dashed lines on the histograms.}
	\label{fig:MCMC_hz}
\end{figure*}

Figure \ref{fig:MCMC_hz} shows the redshift $z > 0.1$ output from the MCMC abundance matching fits. All parameters have low evolution and the SMHM relation evolves only weakly with redshift. For M, $\beta$, and $\gamma$ where the distributions are wide  or close to the prior we have tested wider priors and insignificant change is found.

\section{Fitting the star formation rates}
\label{app:SFR}
The fit to the star-formation rate we derive in Section \ref{subsec:CentMassAcc} is given by the following, Equation \ref{eqn:SFR_DPL},
\begin{equation}
\label{eqn:SFR_DPL}
\begin{split}
SFR(M_*, z) &= 2N(z)\Big[ \Big( \frac{M_*}{M_{n}(z)}\Big) ^{- \alpha(z)} + \Big( \frac{M_*}{M_{n}(z)}\Big)^{\beta(z)} \Big ]^{-1}\\
\log_{10} N(z) &= 10.65 + 0.33z - 0.08z^2\\
\log_{10} M_{n}(z) &= 0.69 + 0.71*z - 0.088z^2\\
\alpha(z) &= 1.0 - 0.022z + 0.009z^2\\
\beta(z) &= 1.8 - 1.0*z - 0.1z^2.
\end{split}
\end{equation}

\section{Comparison of in-situ/ex-situ growth to other models}
\label{app:OtherModels}

\begin{figure*}
	\centering
	\includegraphics[width = \linewidth]{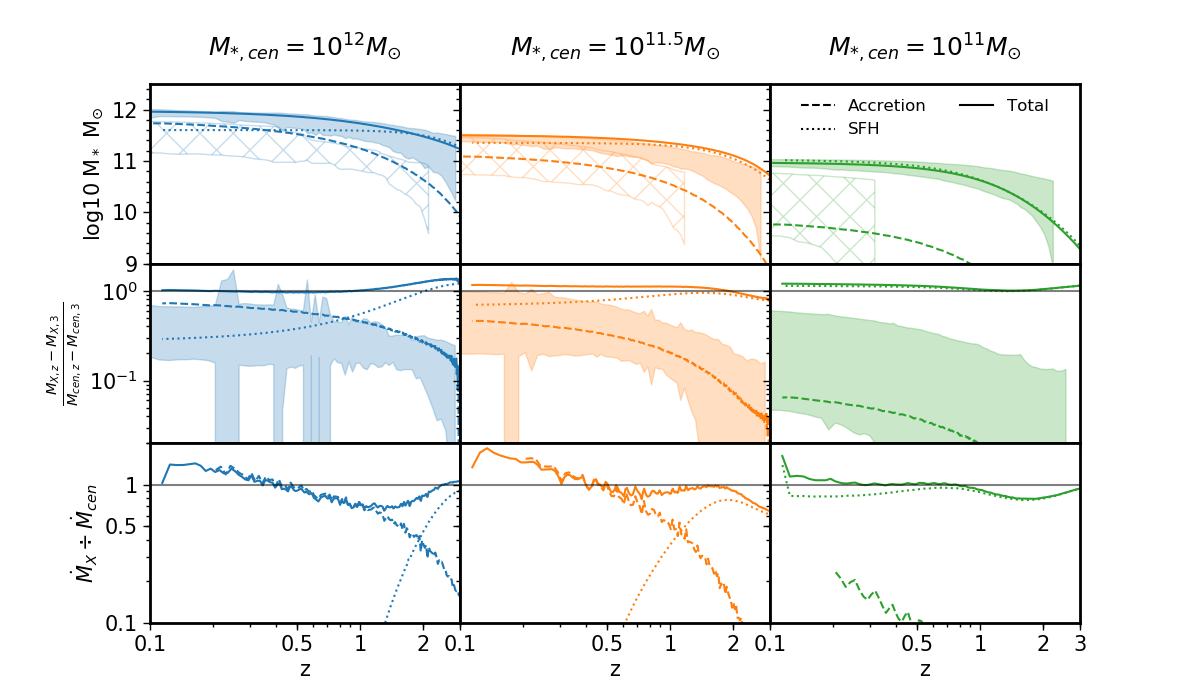}
    \caption{As in Figure \ref{fig:SatelliteAccretion} three `mass tracks' are shown that have central galaxy masses at redshift $z = 0.1$ of $M_{*,cen}$ = $10^{12}$, $10^{11.5}$, and $10^{11}$ $[M_{\odot}]$ in blue orange and green respectively. The satellite galaxy accretion is shown for evolved satellites with a dashed line and the mass from star formation shown with a dotted line. The top panel shows the total mass of the central (solid line) and the total mass gained from accretion or star formation. The middle panel shows the fraction of the total galaxy mass formed from satellite accretion or star formation since redshift $z=3$. The bottom panel shows the ratio of the mass accretion rate from satellite galaxies the star formation rate and the mass growth rate of the central galaxy predicted by abundance matching. In the top panel the shaded regions are galaxies selected from the Illustris simulation the hashed region is then the satellite accretion from Illustris, in the middle panel the shaded region is the ratio of satellite accretion from Illustris. The grey lines in the second and third panel are at unity, the solid lines showing the sum of the other two factors should therefore be close to or on these lines.}
	\label{fig:SatelliteAccretion_ill}
\end{figure*}

 In Figure \ref{fig:SatelliteAccretion_ill} we show the in-situ vs ex-situ growth with the same model as shown in Figure \ref{fig:SatelliteAccretion}, we add to this plot data extracted from the Illustris TNG100 simulation. In Figure \ref{fig:AbnMtch_lz} we see Illustris has a shallower slow mass slope and a steeper high mass slope such that more stellar mass is mapped into haloes of all sizes. We see the change in both of these slopes reflected in the accretion histories, firstly, for the lower mass galaxies (see $log 10 M_{*,cen} = 10^{11}$)  closer to the SMHM knee we find enhanced accretion due to the larger masses from more minor mergers. Secondly the high mass slope is a direct result of the accretion, to support the same merger assembly with the higher mass galaxies in the satellite haloes above the knee where galaxy growth is dominated by the accretion the galaxy growth with halo size must be enhanced.

\begin{figure*}
	\centering
	\includegraphics[width = \linewidth]{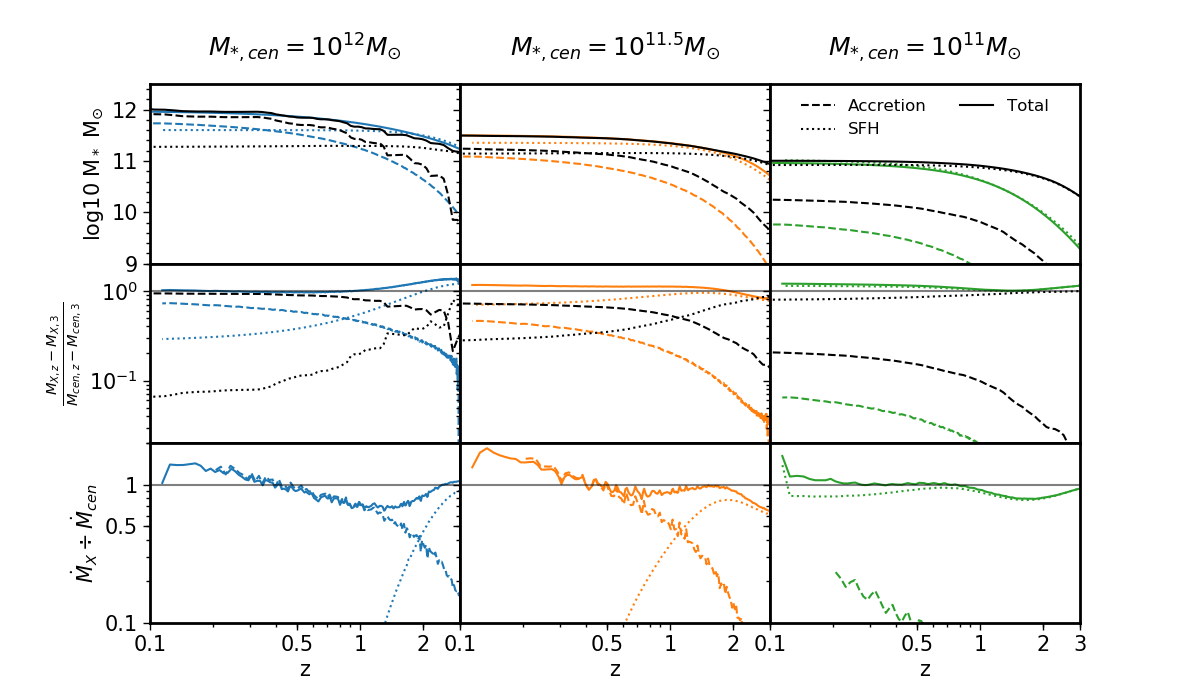}
    \caption{As in Figure \ref{fig:SatelliteAccretion} three `mass tracks' are shown that have central galaxy masses at redshift $z = 0.1$ of $M_{*,cen}$ = $10^{12}$, $10^{11.5}$, and $10^{11}$ $[M_{\odot}]$ in blue orange and green respectively. The satellite galaxy accretion is shown for evolved satellites with a dashed line and the mass from star formation shown with a dotted line. The top panel shows the total mass of the central (solid line) and the total mass gained from accretion or star formation. The middle panel shows the fraction of the total galaxy mass formed from satellite accretion or star formation since redshift $z=3$. The bottom panel shows the ratio of the mass accretion rate from satellite galaxies the star formation rate and the mass growth rate of the central galaxy predicted by abundance matching. In the top and middle rows we add black lines to show the in-situ and ex-situ growth from \textsc{emerge} \citet{Moster2018Emerge10}. The grey lines in the second and third panel are at unity, the solid lines showing the sum of the other two factors should therefore be close to or on these lines.}
	\label{fig:SatelliteAccretion_EMERGE}
\end{figure*}

In Figure \ref{fig:SatelliteAccretion_EMERGE} we show the in-situ vs ex-situ growth with the same model as shown in Figure \ref{fig:SatelliteAccretion}, we add to this plot data from the \textsc{emerge} model from \citet{Moster2018Emerge10} shown as black lines. The solid lines show the total galaxy mass followed back selecting populations by mass at redshift $z = 0.1$. The dotted and dashed lines show the amount of galaxy mass formed in-situ and ex-situ respectively. In all cases \textsc{emerge} predicts satellite accretion becomes the dominant mass growth pathway at higher redshifts then \textsc{steel}. In the third column we see that \textsc{emerge} and \textsc{steel} also disagree about the mass growth history of $log 10 M_{*,cen}$ = 11 galaxies, however, both models agree that the dominant mass growth path of galaxies at this mass are in-situ processes.

\begin{figure*}
	\centering
	\includegraphics[width = \linewidth]{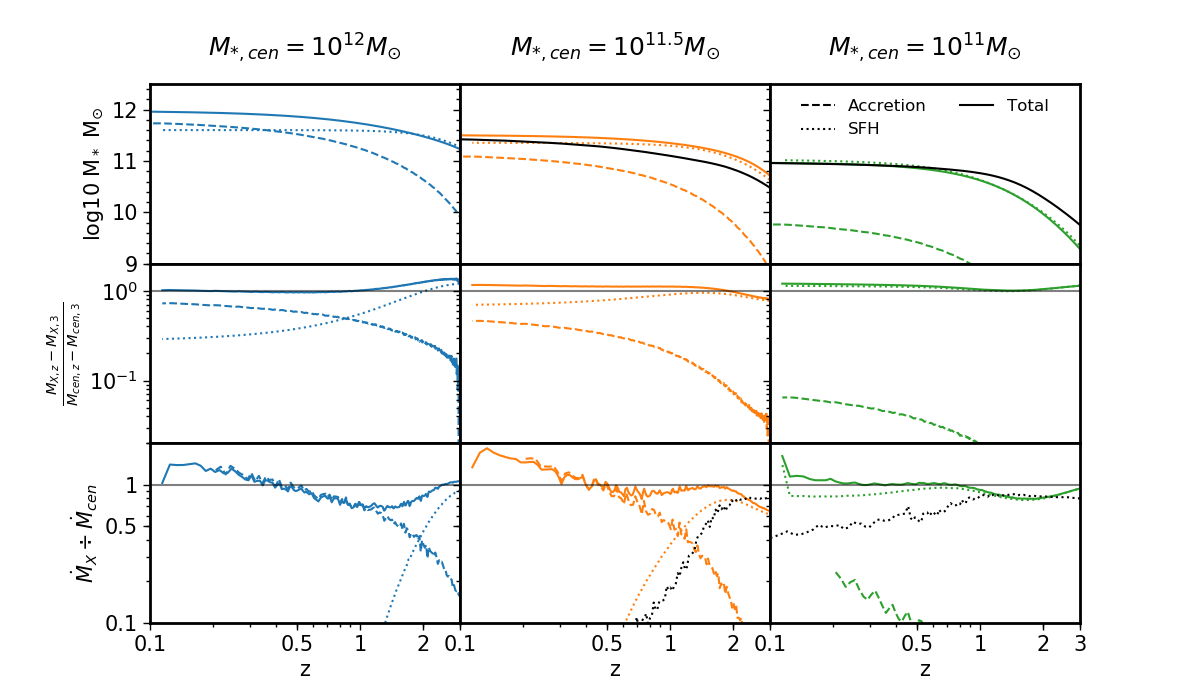}
    \caption{As in Figure \ref{fig:SatelliteAccretion} three `mass tracks' are shown that have central galaxy masses at redshift $z = 0.1$ of $M_{*,cen}$ = $10^{12}$, $10^{11.5}$, and $10^{11}$ $[M_{\odot}]$ in blue orange and green respectively. The satellite galaxy accretion is shown for evolved satellites with a dashed line and the mass from star formation shown with a dotted line. The top panel shows the total mass of the central (solid line) and the total mass gained from accretion or star formation. The middle panel shows the fraction of the total galaxy mass formed from satellite accretion or star formation since redshift $z=3$. The bottom panel shows the ratio of the mass accretion rate from satellite galaxies the star formation rate and the mass growth rate of the central galaxy predicted by abundance matching. In the top and bottom rows, for the $log 10 M_{*,cen}$ = 11.5, and 11, we add black lines to show the central galaxy growth and the star formation rate ratio from \citet{Behroozi2019UniverseMachine:010}. The grey lines in the second and third panel are at unity, the solid lines showing the sum of the other two factors should therefore be close to or on these lines.}
	\label{fig:SatelliteAccretion_UniM}
\end{figure*}

In Figure \ref{fig:SatelliteAccretion_UniM} we show for the $log 10 M_{*,cen}$ = 11.5, and 11 galaxies the central growth and star formation rate ratio from \citet{Behroozi2019UniverseMachine:010}. The central growth is close to that found from \textsc{steel} and the star formation rate transition for  $log 10 M_{*,cen}$ = 11.5 is an excellent match.

\begin{figure*}
	\centering
	\includegraphics[width = \linewidth]{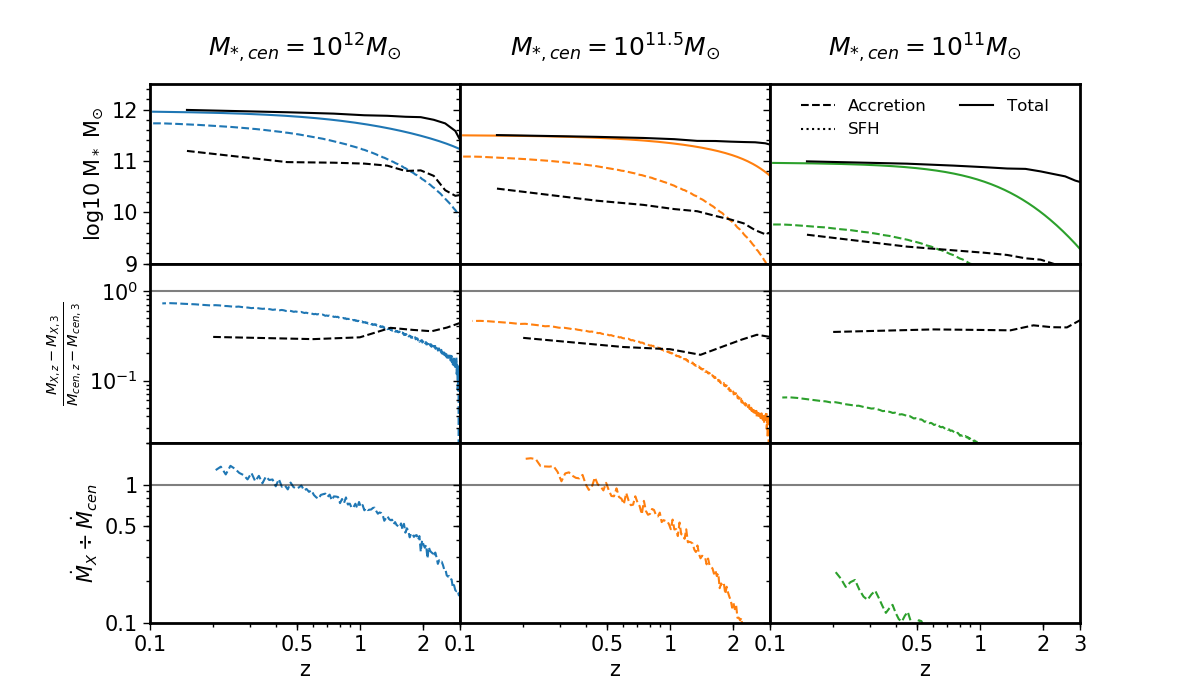}
    \caption{As in Figure \ref{fig:SatelliteAccretion} three `mass tracks' are shown that have central galaxy masses at redshift $z = 0.1$ of $M_{*,cen}$ = $10^{12}$, $10^{11.5}$, and $10^{11}$ $[M_{\odot}]$ in blue orange and green respectively. The satellite galaxy accretion is shown for evolved satellites with a dashed line and the mass from star formation shown with a dotted line. The top panel shows the total mass of the central (solid line) and the total mass gained from accretion or star formation. The middle panel shows the fraction of the total galaxy mass formed from satellite accretion or star formation since redshift $z=3$. The bottom panel shows the ratio of the mass accretion rate from satellite galaxies the star formation rate and the mass growth rate of the central galaxy predicted by abundance matching. In the top and middle rows we add black lines to show the ex-situ growth and central growth from the Semi-Analytic model described in \citet{Menci2014TriggeringInteractions}. The grey lines in the second and third panel are at unity, the solid lines showing the sum of the other two factors should therefore be close to or on these lines.}
	\label{fig:SatelliteAccretion_Menci}
\end{figure*}

\begin{figure}
	\centering
	\includegraphics[width = \linewidth]{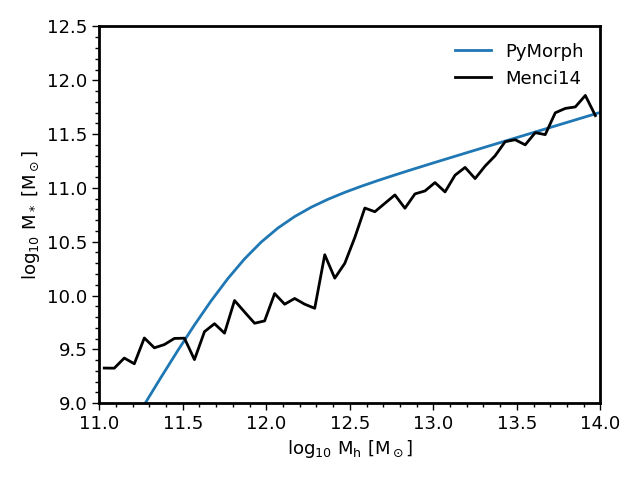}
    \caption{We show the SMHM relationship for PyMorph (blue line, used in this work) and for the Semi-Analytic Model from \citet{Menci2014TriggeringInteractions} (black line).}
	\label{fig:AbundanceMtch_Menci}
\end{figure}

In Figure \ref{fig:SatelliteAccretion_Menci} we show a comparison with the Semi-Analytic model described in \citet{Menci2014TriggeringInteractions}. At all masses the stellar growth is substantially different to \textsc{steel} and the other models shown in this appendix. Furthermore the Semi-Analytic model shows little change in the accreted mass ratio over cosmic time, again this is inconsistent with the findings from \textsc{steel} and the other models presented in this section. We attribute most of the differences seen here to the substantial difference in the SMHM relationship shown in Figure \ref{fig:AbundanceMtch_Menci}.

\section{Halo Merger Rate in the Millennium cosmology.}
\label{app:CosmoAlt}
\begin{figure}
	\centering
	\includegraphics[width = \linewidth]{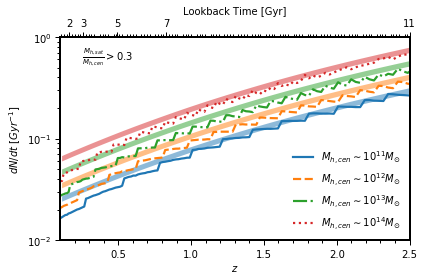}
    \caption{The evolution of merger rate per Gyr at fixed halo mass. Lines are from \textsc{steel} with cosmology altered to that of the analytic fits from \citet{Fakhouri2010TheSimulations} (shaded bands). Halo masses shown are $M_{h,cen}: 10^{11}, 10^{12}, 10^{13}, 10^{14} M_{\odot} h^{-1}$ as labelled.}
	\label{fig:HaloMergerRate_Mill}
\end{figure}

\RevHighlight{}{The statistical dark matter accretion history used in \textsc{steel} is cosmologically flexible}\footnote{\RevHighlight{}{The eventual goal of \textsc{steel} is to be cosmologically independent such that any cosmology can be used.}}. \RevHighlight{}{The current prescriptions for the halo growth histories and dark matter substructure \citep{vandenBosch2014ComingWells, Jiang2016StatisticsFunctions} can be used for $\Lambda$CDM models that are within a factor two of current constraints.} \RevHighlight{}{Using \textsc{collosus} \citep{Diemer2017COLOSSUS:Halos} we set the cosmology used in \textsc{steel} to that of the Millennium simulation ($\Omega_{m}$ = 0.25, $\Omega_{b}$ = 0.045, $\Omega_{\Lambda}$ = 0.75, h = 0.73). The statistical dark matter accretion history is then recalculated. Figure \ref{fig:HaloMergerRate_Mill} shows the halo merger rate from \textsc{steel} from this alternative accretion history. The halo merger rate tracks from \citet{Fakhouri2010DarkDependence} shown in Figures \ref{fig:HaloMergerRate} \& \ref{fig:HaloMergerRate_Mill} are calculated using the Millennium simulation cosmology. The deviation between the merger rate of \textsc{steel} and that of \citet{Fakhouri2010DarkDependence} still remains despite the change of cosmology, as discussed in Section \ref{subsec:StatMergeHist} this is an effect of the algorithms used to build halo merger trees.}

\bsp	
\label{lastpage}
\end{document}